\documentclass[runningheads]{llncs}

\usepackage[dvipsnames]{xcolor}

\usepackage{etoolbox}
\newtoggle{longversion}
\toggletrue{longversion}
\newtoggle{isitlongversion}
\usepackage{url}
\usepackage{enumitem}

\usepackage{amsmath,amssymb,amsfonts,xspace}
\usepackage{mathtools,bm,bbm}
\usepackage{thm-restate}

\usepackage{hyperref}
\hypersetup{
  colorlinks   = true, 
  urlcolor     = black!50!red, 
  linkcolor    = black!50!brown, 
  citecolor   = black!50!brown 
}
\usepackage[capitalize]{cleveref}

\usepackage{verbatim}
\usepackage{mdwtab}
\usepackage{subfigure}
\usepackage{authblk}
\usepackage{hhline}
\usepackage{multirow}
\usepackage{pdfpages}
\usepackage{cite}
\usepackage{algorithmic}
\usepackage{textcomp}
\usepackage{soul}
\usepackage{booktabs}

\usepackage{pgfplots}
\pgfplotsset{compat=1.15}
\usepackage{color,graphicx,tikz}
\usetikzlibrary{circuits.ee.IEC}
\usetikzlibrary{arrows,shapes.geometric}

\mathtoolsset{showonlyrefs} 

\tikzset{
  declare function={
    atan3(\a,\b)=ifthenelse(atan2(0,1)==90, atan2(\a,\b), atan2(\b,\a));},
  kinky cross radius/.initial=+.125cm,
  @kinky cross/.initial=+, kinky crosses/.is choice,
  kinky crosses/left/.style={@kinky cross=-},kinky crosses/right/.style={@kinky cross=+},
  kinky cross/.style args={(#1)--(#2)}{
    to path={
      let \p{@kc@}=($(\tikztotarget)-(\tikztostart)$),
          \n{@kc@}={atan3(\p{@kc@})+180} in
      -- ($(intersection of \tikztostart--{\tikztotarget} and #1--#2)!%
             \pgfkeysvalueof{/tikz/kinky cross radius}!(\tikztostart)$)
      arc [ radius     =\pgfkeysvalueof{/tikz/kinky cross radius},
            start angle=\n{@kc@},
            delta angle=\pgfkeysvalueof{/tikz/@kinky cross}180 ]
      -- (\tikztotarget)}}}

\newcommand{\pink}{\textcolor{black}}
\newcommand{\cA}{\mathcal{A}}
\newcommand{\cC}{\mathcal{C}}
\newcommand{\cE}{\mathcal{E}}

\newcommand{\cJ}{\mathcal{J}}
\newcommand{\cM}{\mathcal{M}}
\newcommand{\cP}{\mathcal{P}}
\newcommand{\cS}{\mathcal{S}}

\newcommand{\cX}{\mathcal{X}}
\newcommand{\cY}{\mathcal{Y}}
\newcommand{\cZ}{\mathcal{Z}}
\newcommand{\cQ}{\mathcal{Q}}

\newcommand{\defineqq}{\ensuremath{\stackrel{\textup{\tiny def}}{=}}}

\usetikzlibrary{shapes,arrows,positioning,decorations.pathreplacing,calc}

\def\nummsg{\mbox{$N$}} 

\newcommand{\black}{\color{black}}

\newcommand{\ina}[1]{\left<#1\right>}
\newcommand{\inb}[1]{\left\{#1\right\}}
\newcommand{\inp}[1]{\left(#1\right)}

\def\deterministic{\ensuremath{\mathrm{deterministic}}}
\def\random{\ensuremath{\mathrm{random}}}

\newcommand{\mo}{\ensuremath{m_{\one}}\xspace}

\newcommand\independent{\protect\mathpalette{\protect\independenT}{\perp}}
\def\independenT#1#2{\mathrel{\rlap{$#1#2$}\mkern2mu{#1#2}}}

\def\one{\ensuremath{\mathsf{A}}\xspace} 
\def\two{\ensuremath{\mathsf{B}}\xspace} 

\newcommand{\bmac}{\ensuremath{\text{Byzantine-MAC}}\xspace}

\newcommand{\red}[1]{{\textcolor{red}{#1}}}

\DeclareMathOperator{\Tr}{Tr}
\begin{document}

\title{Communication with Adversarial Users:
Byzantine Models in Multi-User Information Theory}

\titlerunning{Communication with Adversarial Users}

\author{
Minglai Cai\inst{1}
\and
Christian Deppe\inst{2}
\and
Neha Sangwan\inst{3}
}

\authorrunning{M. Cai, C. Deppe, N. Sangwan}

\institute{
Department of Mathematics and Computer Science,
Universit\"at zu K\"oln, Germany\\
\email{mcai1@uni-koeln.de}
\and
Institute for Communications Technology (IfN),
Technische Universit\"at Braunschweig, Germany\\
\email{christian.deppe@tu-bs.de}
\and
University of California San Diego (UCSD), USA\\
\email{nehasangwan010@gmail.com}
}

\maketitle

\begin{abstract}
    We consider communication systems  operating in environments where some participating users cannot be trusted. This motivates the study of Byzantine communication models, in which  users may behave adversarially. In this survey, we provide a unified information-theoretic perspective on communication with adversarial users in multi-user settings. We review recent advances in Byzantine multiple access channels, including reliable communication, adversary identification, and distributed function computation. Furthermore, we discuss extensions to classical-quantum channels and highlight fundamental differences between classical and quantum models. Finally, we outline key open problems and future research directions at the intersection of communication, computation, and security.

\end{abstract}


\section{Introduction}
\label{sec:intro}
    Modern communication systems, such as wireless Internet-of-Things (IoT) networks and large-scale distributed sensing architectures, consist of devices with heterogeneous capabilities and varying levels of trust~\cite{8232533,ammar2018internet}. Since these devices share a common communication medium, a compromised device may attempt to disrupt the communication process from within the system. This raises fundamental questions about reliable communication in the presence of \emph{adversarial users}, i.e., users who do not follow the prescribed protocol and may act maliciously.
    \color{black}{A natural information-theoretic framework for studying such communication systems is the \emph{Byzantine Multiple Access Channel} (Byzantine MAC), which extends the classical \emph{Multiple Access Channel} (MAC) to accommodate adversarial users. A MAC models communication between multiple senders and a single receiver over a shared communication medium. In Byzantine MAC,} one or more of the users may deviate arbitrarily from the  given protocol. The receiver, however, is unaware of whether any user is adversarial and does not know the identity of such users. 
    \black{A distinguishing feature of this model is that the adversary is not an external entity but rather one or more of the legitimate participants in the communication system.}
    
    This Byzantine perspective differs substantially from classical adversarial models in information theory. In the wiretap channel~\cite{Wyner75}, the adversary is passive and observes a noisy version of the communication, while the goal is to ensure reliability and secrecy. In arbitrarily varying channels (AVCs)~\cite{BlackwellBTAMS60,survey}, the adversary is active and controls the channel state in order to jam communication. The multi-user extension, arbitrarily varying multiple access channels (AV-MACs), has been studied extensively~\cite{Jahn81,Gubner90,Ahl/Ca,WieseB13,PeregS19,BeemerCNS20}. In all these models, however, the adversary is external to the system. In contrast, Byzantine models capture \emph{internal adversaries}, i.e., one of the ``legitimate'' users that may behave maliciously (cf. \cref{intext}).
    
    Over the past years, a sequence of works has  developed an information-theoretic understanding of such systems. \emph{Reliable communication} in Byzantine MACs has been studied in~\cite{NehaBDP19}. The objective is to decode all trustworthy users’ messages despite adversarial interference. This line of work reveals that the presence of adversarial users fundamentally alters the structure of the capacity region and introduces new feasibility conditions that do not arise in conventional AV-MACs. 
    \black{The notion of \emph{authenticated communication} in Byzantine MACs was introduced in~\cite{NehaBDP19}. In this model, the decoder is required to either reliably decode the messages of the trustworthy users or detect the presence of adversarial interference. Thus, unlike in the reliable communication model, the decoder is allowed to forgo decoding when an adversarial attack occurs. An intermediate model, termed \emph{communication with adversary identification}, was introduced in~\cite{NehaBDP21}. In this model, the decoder may declare adversarial interference instead of decoding the trustworthy users' messages, but must additionally identify the adversarial user (in two user Byzantine MAC) when doing so.} A detailed comparison between all these models  is given in~\cite{Sa/Ba/De/Pr2}.
    More recently, Byzantine distributed function computation has been investigated in~\cite{AnilkumarSPNP25}, where the objective shifts from message transmission to robust recovery of functions of distributed data in the presence of adversarial users.

    A unifying phenomenon underlying all these models is the presence of competing
    adversarial explanations for the same received observation. In the \black{classical communication}
    setting, an adversarial user may generate channel input that imitate plausible
    honest behavior, thereby creating ambiguity at the decoder. In distributed
    function computation, different \black{set of adversarial users} may induce statistically
    indistinguishable observations while leading to different function values.
    Consequently, reliable
    communication or computation is fundamentally possible only for quantities that
    remain invariant under all adversarial scenarios that are indistinguishable to
    the decoder.
    
    While these developments already exhibit rich and fundamentally new phenomena in the classical setting, the situation becomes even more intricate when moving to quantum communication systems. In the quantum scenario, many new challenges arise that have no direct classical analogue. Most importantly, quantum measurements are inherently invasive: every measurement may change the quantum state. As a consequence, not only the input of an adversarial sender may disrupt the communication process, but also previously applied decoding operations can influence subsequent decoding steps. This creates a fundamentally different interaction between encoding, decoding, and adversarial behavior compared to the classical systems.
    
    Furthermore, the role of randomness in coding theory changes significantly between  classical and  quantum setting. In the classical setting, it is well known that the capacity of arbitrarily varying channels can be achieved by deterministic codes under suitable conditions~\cite{Cs/Na}. Similarly, for arbitrarily varying multiple access channels, deterministic coding schemes have been constructed~\cite{Ahl/Ca}. In contrast, for quantum arbitrarily varying channels, only random coding techniques are currently available~\cite{Ahl/Bli,Da/Wa/Ha}. This indicates a fundamental limitation in transferring classical derandomization and robustness techniques to the quantum domain.
    
    These differences have profound implications for Byzantine communication.
    Many classical techniques developed for Byzantine MACs, including deterministic
    verification procedures and identification-based decoding methods, do not
    directly extend to quantum channels. In particular, an unknown quantum state generally cannot be verified with certainty: even an invalid quantum state may pass a decoding
    measurement with non-vanishing probability. Consequently, fundamentally new
    coding and decoding techniques are required for quantum Byzantine communication.

    \paragraph*{Organization.} The remainder of this paper is organized as follows.
\Cref{sec:taxonomy} introduces a taxonomy of Byzantine communication models and decoding guarantees.
\Cref{sec:unify} presents unifying principles and conceptual insights underlying the considered models.
\Cref{sec:connections} discusses connections to related research areas, including arbitrarily varying channels, distributed computation, cryptography, and semantic communication.
\Cref{sec:classical_bmac} introduces the classical Byzantine multiple access channel model.
Section \ref{rc}  describes the 
 \black{notion of } \emph{reliable communication}, where the decoder is required to recover the messages of all trustworthy users with high probability, regardless of the behavior of the adversarial users~\cite{Sa/Ba/De/Pr}. In this setting, no guarantees are required for the messages of adversarial users.
In Section \ref{aut}, we consider a related but weaker notion, the \emph{authenticated communication}, where the decoder must either output the correct messages or declare an error, ensuring that adversarial users cannot induce undetected decoding errors.
A stronger objective, introduced in Section \ref{iden}, is  \emph{communication with adversary identification}, where the decoder must additionally identify the adversarial user when it cannot reliably decode all users’ messages ~\cite{NehaBDP21}. This introduces an additional layer of robustness and leads to new feasibility conditions.
\Cref{sec:distributed_computation} discusses \emph{Byzantine distributed function computation} and robust task-oriented communication. The goal of distributed function computation \black{is to either compute a function of the users' data reliably or identify} adversarial participants~\cite{AnilkumarSPNP25}.
In Section \ref{sec:quantum_bmac}, we  explore a quantum counterpart of  Byzantine multiple access channels. Specifically, we investigate a classical-quantum multiple access channel in which one of the transmitters may act adversarially.
\black{This requires developing novel coding approach to overcome new challenges arising in quantum systems.} 
\cref{sec:open_problems} outlines open problems and future research directions. 
\black{We finally conclude the survey in \cref{sec:conclusion}.}


\section{A Taxonomy of Byzantine Communication Models}
\label{sec:taxonomy}

    In this section, we introduce a unified taxonomy for communication systems with adversarial users. The goal is to clarify the different modeling choices that arise in the literature and to provide a common framework that allows us to relate the various results discussed in this survey.
    


\subsection{External vs.\ Internal Adversaries}
\label{intext}
    A first fundamental distinction is between \emph{external} and \emph{internal} adversaries. In classical models, such as the arbitrarily varying channel (AVC)~\cite{BlackwellBTAMS60,Cs/Na} and its multi-user extensions~\cite{Jahn81,Gubner90,Ahl/Ca}, the adversary is external to the communication system and acts by controlling the channel state. The trustworthy users follow the protocol, and the adversary influences the communication only indirectly through the channel.
    
    In contrast, Byzantine models consider \emph{internal adversaries}, i.e., adversaries that are themselves trustworthy users of the system. Such users have full access to the protocol and may deviate arbitrarily from it. This seemingly minor distinction leads to a rich set of modeling dimensions ~\cite{Sa/Ba/De/Pr,Sa/Ba/De/Pr2}, where the adversarial behavior is modeled at the level of the users rather than the channel.

    A natural connection exists between Byzantine communication and arbitrarily
    varying channels (AVCs). In both models, communication must remain reliable
    despite adversarial behavior.
    It can be observed that several conceptual parallels emerge. The role of
    symmetrizability in AVCs is closely related to \black{ the notions of symmetrizability, spoofability and overwritability} in Byzantine
    systems. In both cases, the adversary attempts to create statistically
    indistinguishable communication scenarios. Furthermore, many proof techniques,
    including robustification arguments, random coding, and worst-case analysis,
    have close analogues across both frameworks.
    
    At the same time, Byzantine communication introduces substantially new
    difficulties because the adversary participates directly in the communication
    protocol. This internal structure creates new interactions between coding,
    verification, and decoding that do not appear in standard AVC models.

\subsection{Jammer with Partial Knowledge vs. 
Jammer with Full Knowledge}
\label{Akac}
    A second important dimension concerns the knowledge and capabilities of the adversary.

 In the AVC model, a \emph{jammer with partial knowledge}
knows the overall system design, i.e., the used encoder and
decoder.
\textcolor{black}{However, the jammer does not know the
actual transmitted message.}
This reflects \textcolor{black}{the situation in}
public (quantum) communication networks,
\textcolor{black}{where} the used coding schemes are
standardized and publicly available.
A \emph{jammer with full knowledge} is more powerful
and is further aware of the actual transmitted message.
The scenario of a jammer with partial knowledge
\textcolor{black}{is naturally associated with}
an AVC under the average error criterion,
while \textcolor{black}{a fully informed jammer may adapt
its strategy to the transmitted message, thereby capturing
the worst-case behavior underlying the maximal-error
criterion (cf. \cite{Bo/Ca/Ca}).}

\textcolor{black}{There is a close connection between
the maximal-error capacity of AVCs and Shannon's
zero-error capacity problem, introduced in 1956
\cite{Sha}. In particular, Ahlswede \cite{Ahl0}
showed that the zero-error capacity of a discrete
memoryless channel equals the maximal-error capacity
of a suitably constructed $0$--$1$ AVC.
Consequently, a general solution to the maximal-error
AVC capacity problem would also solve Shannon's
longstanding zero-error capacity problem.}

\cite{Bo/Ca/Ca} shows that there is no Ahlswede
dichotomy for AVCs under the maximal-error criterion:
It may occur that the deterministic capacity of an AVC
under the maximal-error criterion is
\textcolor{black}{strictly positive but smaller than
its random-code capacity.}

Interestingly, \cite{Bo/No} shows that
the deterministic capacities of
\textcolor{black}{finite quantum-to-quantum AVCs
for classical message transmission}
under \textcolor{black}{the maximal-error and
average-error criteria coincide, since mixed quantum
input states allow randomized encoding even within
the deterministic-code framework}
(cf. Section \ref{ClvsQu} for the distinction
between deterministic and random coding).
    
 By virtue of the facts above,    we consider 
adversarial users
with partial knowledge
in the Byzantine MAC models  in this survey.

    
    Another key aspect is whether adversarial users can \emph{collude}. In the simplest Byzantine MAC models~\cite{Sa/Ba/De/Pr}, at most one user is assumed to be adversarial. More general formulations allow multiple adversarial users or even adversarial subsets, leading to the notion of an adversary structure, which plays a central role in distributed computation settings~\cite{AnilkumarSPNP25}.


    
    
    

\subsection{Symmetrizability vs. Spoofability vs. Overwritability}
\label{SyandSp}
    One of the central concepts in adversarial information theory is
    symmetrizability. In arbitrarily varying channels (AVCs) and arbitrarily varying
    multiple access channels (AV-MACs), symmetrizability characterizes situations
    in which an external adversary can confuse different channel inputs so that the
    receiver cannot reliably distinguish between them. In such cases, reliable
    communication with deterministic codes becomes impossible.
    
    A closely related phenomenon appears in Byzantine communication models.
    The main challenge here is that, since the adversary is not external to the channel, but rather
    coincides with one of the trustworthy users, it does not
    control the channel state directly. Instead, it manipulates the transmitted
    inputs themselves. \color{black}{This leads to notions of symmetrizability and spoofability in Byzantine communication models. While symmetrizability describes the possibility of confusing different transmitted messages, spoofability captures the stronger requirement that an adversarial user can create ambiguity about both the transmitted messages and the identity of the adversarial user.}
    
    \black
    In classical Byzantine MACs, spoofability arises whenever an adversarial user can
    choose inputs such that the resulting channel output becomes statistically
    indistinguishable from that of an honest transmission scenario (cf. Section \ref{SPOOF}). In this case,
    the decoder cannot uniquely determine which user behaved maliciously. The ambiguity illustrated in Fig.~1 provides a concrete example of how
    indistinguishable adversarial explanations arise in Byzantine communication.
    
    More generally, the converse proofs for Byzantine communication problems follow
    a common indistinguishability principle. If two adversarial scenarios induce the
    same distribution at the decoder, then no decoder can reliably distinguish
    between them. Therefore, any reliably recoverable quantity must coincide under
    all such indistinguishable scenarios. In distributed function computation, this
    principle leads directly to the $s$-viability condition. In Byzantine MACs, it
    determines which rate pairs remain achievable despite adversarial interference.
    
    In the quantum setting, the notion of spoofability becomes significantly more
    subtle. Since quantum states cannot generally be verified perfectly, an
    adversarial user may generate states that pass the decoding measurement of a
    legitimate codeword with non-vanishing probability. Moreover, measurements are
    destructive and may alter the channel output seen by subsequent decoding
    operations. Consequently, spoofability in quantum Byzantine communication is
    not only a property of the transmitted states, but also of the measurement
    strategy itself.
    
    This observation reveals an important conceptual connection between Byzantine
    communication and classical symmetrizability theory. In both settings, reliable
    communication becomes impossible whenever adversarial actions can create
    indistinguishable observations on the decoder. However, Byzantine models extend
    this principle from channel-state manipulation to strategic manipulation by
    internal users.
    
    The resulting viewpoint suggests that symmetrizability and spoofability are
    different manifestations of a common phenomenon: the inability of the decoder to
    distinguish between competing explanations of the received observation. This
    perspective unifies classical AVCs, Byzantine MACs, distributed function
    computation, and quantum Byzantine communication within a single structural
    framework.

An even stronger condition is the overwritable condition (cf. Section \ref{OVERW}).
Overwritability
captures the stronger condition wherein the adversary 
can not only
 make one message
 indistinguishable, but
can also
completely replace the legitimate message with a counterfeit,
without giving away its presence.
In other words, the adversary can  inject a
spurious message which will appear equally plausible to the
decoder as if there is no attack at all.
Notice that all overwritable Byzantine-MACs are spoofable and all spoofable Byzantine-MACs are symmetrizable. Furthermore,
both  inclusions are strict (cf. \cite{Sa/Ba/De/Pr}).

\subsection{Strong Deterministic vs.\ Weak Deterministic and Randomized Coding}
\label{ClvsQu}
    Another important dimension is the type of coding strategies that are allowed (cf. Section \ref{randet}).

For classical arbitrarily varying channels and AV-MACs, randomized coding can
    often overcome symmetrizability effects created by an adversarial jammer.
The arbitrarily varying channel was first introduced
 in \cite{Bl/Br/Th2}. 
 \cite{Ahl1}  showed a surprising result
that is now known as the Ahlswede Dichotomy:
The 
capacity of an arbitrarily varying channel is either zero or  equal to
 its shared randomness assisted capacity. 
The idea to show the Ahlswede Dichotomy is to build   a two-part code word, the first part is a short deterministic
code word,
creating the common randomness for the sender and the legal receiver, and the second part is a randomized code word, transmitting the actual message to the legal
receiver.
\cite{Cs/Na} provides the condition when exactly the deterministic capacity is positive
 by  constructing
 a one-part deterministic code.
We denote the two-part code concept by weak deterministic
code and the one-part code concept by strong deterministic code.


{\begin{center}\begin{figure}
\begin{minipage}{0.55\linewidth}
        \centering
\includegraphics[width=0.95\linewidth]{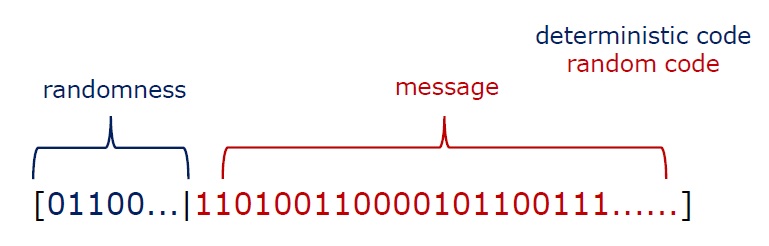}
\caption{Weak deterministic coding concept. Two-part code word:
a deterministic code
to create shared randomness;
a  randomness assisted code
to transform the message
}
 \end{minipage}
  \hspace{.05\linewidth}
\begin{minipage}{0.35\linewidth}
        \centering
\includegraphics[width=0.95\linewidth]{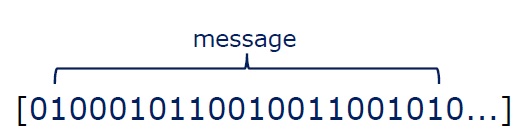}
\caption{Strong deterministic coding concept. One-part deterministic secure code. 
Technically difficult to extend  to quantum channels}
 \end{minipage}
\label{str}
\end{figure}\end{center}}

    In Byzantine MACs, similar phenomena arise. Randomized coding schemes, often based on shared randomness between users and the decoder, can achieve larger rate regions~\cite{Sa/Ba/De/Pr}. Deterministic coding, on the other hand, requires more stringent structural conditions and is closely related to generalized notions of symmetrizability.


    A key difference between classical Byzantine communication and classical
    AV-MACs is that the adversary in Byzantine models is internal to the system.
    Consequently, the adversary has protocol-level knowledge and can strategically
 imitate honest behavior. Randomization therefore not only protects against
    worst-case channel effects, but also helps conceal structural information about
    the coding scheme itself.
    
    In distributed function computation, randomness plays a less direct role. Since
    the users may transmit their observations uncoded, the robustness mechanism is
    primarily based on statistical consistency and viability conditions rather than
    on random codebook generation. Nevertheless, the underlying proofs still rely
    heavily on probabilistic arguments, including type concentration and typicality
    methods.

In the quantum setting, the distinction
between deterministic and random coding
becomes even more pronounced. For general classical-quantum arbitrarily varying channels, only random coding techniques and  weak deterministic coding techniques are currently known to achieve capacity~\cite{Ahl/Bli,Da/Wa/Ha}. This limitation carries over to Byzantine models and highlights the need for fundamentally new coding techniques in the quantum domain~\cite{CaiQuantumByzantineMAC}.
 The existing capacity
 results  are fundamentally based on random coding techniques. This limitation
    extends naturally to quantum Byzantine multiple access channels.
    
    The reason is deeply connected to the structure of quantum measurements. In
    classical systems, strong deterministic decoding regions can often be designed such
    that adversarial manipulations become distinguishable. In quantum systems,
    however, decoding is performed through POVM measurements that inherently
    introduce probabilistic behavior and measurement disturbance. Consequently,
    classical derandomization techniques do not directly extend to the quantum
    domain.
    
    Thus, the role of randomness evolves across the models considered in this
    survey. In classical adversarial communication, randomness primarily combats
    symmetrizability and spoofing attacks. In quantum Byzantine communication, it
    becomes an intrinsic component of reliable communication itself. This highlights
    a fundamental conceptual difference between classical and quantum robustness
    theory.
    
    Taken together, the considered models suggest that Byzantine communication
    problems can be viewed as inference problems under adversarial ambiguity.
    The decoder must recover only those quantities that remain invariant under
    all statistically or operationally indistinguishable adversarial explanations.
    This viewpoint provides a unifying perspective across classical communication,
    distributed computation, and quantum information processing.



    
    \medskip
    
    The taxonomy introduced in this section provides a conceptual framework for the remainder of the survey. In the following sections, we explore these different dimensions in detail, starting with classical Byzantine multiple access channels and their capacity characterizations.
\section{Unifying Principles and Insights}
\label{sec:unify}
    The models discussed in the survey differ in their technical
    formulation. But they also share several common structural principles. In all cases,
    the central difficulty is that the adversary  acts as one of the legitimate participants. This internal nature of
    the adversary changes the role of decoding: the receiver must not only recover
    messages or function values, but also reason about which users may have behaved
    maliciously.
    

    A first common feature is the presence of several competing explanations for the
    same received observation. In a classical Byzantine multiple access channel, the
    same channel output may be consistent with different combinations of honest and
    adversarial behavior. In distributed function computation, the same reported
    source sequences may be ``explained" by different malicious subsets. In the quantum
    setting, the situation is even more delicate, since measurements themselves may
    change the received state.
    
    This leads to a general principle: reliable recovery is possible only for those
    quantities that are invariant under all indistinguishable adversarial
    explanations. In communication problems, this invariance is reflected in the
    structure of the achievable rate region. In distributed function computation, it
    appears explicitly through the $s$-viability condition. In quantum Byzantine
    communication, the same idea persists, but indistinguishability must be
    understood in terms of quantum measurements and state discrimination.
    
    A second common feature is the separation between trustworthy-user recovery and
    adversary treatment. The decoder is not required to decode the adversarial user's
    message or reconstruct the adversarial user's source sequence. Instead, it must
    protect the trustworthy users: either their information is recovered correctly, or a
    malicious participant is identified. This distinction is essential for Byzantine
    models and separates them from standard multiple access communication.
    
    A third recurring feature is the role of side information. In the classical BMAC,
    the channel output provides the only evidence about the users' behavior. In
    distributed function computation, the decoder may additionally have side
    information $Y^n$, which can help verify the consistency of reported source
    sequences. In the quantum setting, however, the available information is limited
    by the measurement process itself. Thus, side information and verification are
    central themes across all models, but their operational meaning depends on the
    underlying communication system.

\section{Connections to Other Areas}
\label{sec:connections}

    The study of Byzantine communication models is closely connected to several
    other research areas, including arbitrarily varying channels, distributed
    computation, cryptography, network coding, and task-oriented communication.
    Although these areas were historically developed largely independently, they
    share many structural similarities with the models considered in this survey.
    

    
    
    
\subsection{Connections to Distributed Computation and Learning}

    Byzantine distributed function computation is closely related to distributed
    computation and distributed learning systems. In many modern architectures,
    multiple agents collaboratively provide local observations or intermediate
    results to a central processor. Examples include federated learning, distributed
    sensor networks, edge computing, and cooperative autonomous systems (\cite{Dist1,Dist2}).
    
    In such systems, robustness against malicious participants is essential.
    Byzantine users may intentionally manipulate local updates, falsify sensor
    measurements, or inject inconsistent information into the computation process.
    The information-theoretic framework considered in this survey provides a
    fundamental characterization of which functions remain reliably computable under
    such adversarial behavior.
    
    The viability condition obtained for Byzantine distributed function computation
    is particularly relevant in this context. It formalizes the idea that only those
    quantities that remain invariant under all statistically indistinguishable
    adversarial manipulations can be computed robustly (cf. the discussion in \Cref{distdis} and \Cref{leandad}).

\subsection{Connections to Cryptography and Authentication}

    There are also strong connections between Byzantine communication and
    information-theoretic cryptography(\cite{Hi/Ma, Fi/Hi}). In both areas, the goal is to protect
    communication systems against malicious participants without relying on
    computational hardness assumptions.
    
    In particular, Byzantine communication models are closely related to
    authentication problems. The decoder must ensure that adversarial users cannot
    induce incorrect outputs while remaining undetected. This resembles the goals of
    message authentication and secure broadcast protocols in cryptography (cf. the discussion in \Cref{autdis}).
    
    Moreover, distributed function computation with adversarial users is related to
    secret sharing, secure multiparty computation, and Byzantine agreement. The
    adversary structures considered in distributed computation correspond naturally
    to the adversary models used in cryptographic protocols.
    
    An important difference, however, is that the present framework focuses on
    fundamental information-theoretic limits rather than protocol design. The
    results therefore characterize which tasks are theoretically possible under
    statistical uncertainty and adversarial manipulation.

\subsection{Connections to Task-Oriented and Semantic Communication}

    The considered models also have strong conceptual connections to task-oriented
    and semantic communication. In classical communication theory, the objective is
    typically the reliable reconstruction of transmitted messages. In contrast,
    task-oriented communication focuses on recovering only the information relevant
    for a downstream task.
    
    Byzantine distributed function computation naturally fits into this paradigm.
    The decoder is interested only in a prescribed function of the distributed data,
    rather than the complete observations themselves. Consequently, robustness can
    sometimes be maintained even when full source recovery is impossible.
    
    This perspective is particularly relevant for future communication systems,
    including integrated sensing and communication (ISAC), distributed autonomous
    control, and semantic communication architectures. In such systems, robustness
    against malicious participants may depend critically on the task that the
    system is designed to accomplish (\cite{Go/Mi2,Go/Mi}).

\subsection{Connections to Quantum Information Theory}

    Finally, the quantum extensions discussed in this survey connect Byzantine
    communication to several fundamental topics in quantum information theory (cf. Section \ref{sec:quantum_bmac}),
    including quantum hypothesis testing, state discrimination, arbitrarily varying
    quantum channels, and quantum network communication.
    
    The quantum setting reveals that many classical robustness concepts depend
    implicitly on properties unique to classical information, such as non-invasive
    observation and deterministic verification. Once measurements become invasive
    and non-commutativity enters the model, entirely new challenges arise.
    
    As a result, Byzantine quantum communication lies at the intersection of
    multi-user quantum Shannon theory, adversarial communication, and quantum
    cryptography. The currently available results represent only first steps toward
    a more comprehensive theory of robust quantum network communication (cf.  \Cref{robdis}).

\section{Classical Byzantine Multiple Access Channels}
\label{sec:classical_bmac}

    In this section, we introduce the classical Byzantine multiple access channel (BMAC), which forms the central model for communication with adversarial users. We follow the framework developed in~\cite{Sa/Ba/De/Pr,Sa/Ba/De/Pr2} and present the main definitions and performance criteria.

\subsection{Notation}

    We briefly summarize the notation used throughout the paper. Random variables are denoted by capital letters such as $X$ and $Y$, and their alphabets by calligraphic letters such as $\cX$ and $\cY$. Vectors of length $n$ are denoted by boldface symbols such as $x^n \in \cX^n$.
    For an alphabet $\cX$, we denote by $\cP^n_{\cX}$ the set of empirical distributions (types) of sequences in $\cX^n$. For $P_X \in \cP^n_{\cX}$, the corresponding type class is denoted by $T^n_X$. For a joint type $P_{XY}$, we define the conditional type class
    \[
    T^n_{Y|X}(x^n) \defineqq \{y^n : (x^n,y^n) \in T^n_{XY}\}.
    \]
    
    For a conditional distribution $P_{Y|X}$, the corresponding memoryless channel is denoted by $P^n_{Y|X}$.

\subsection{System Model}

    We consider a $K$-user discrete memoryless multiple access channel (MAC) $W$ with finite input alphabets  $\mathcal{X}_1,\mathcal{X}_2,\ldots, \mathcal{X}_k$, and output alphabet ${\mathcal Y}$.  The channel is used over $n$ time steps and is assumed to be memoryless.

    In contrast to the classical MAC, we allow for the presence of an adversarial \black{users}. Specifically, \black{a set of} users may deviate arbitrarily from the encoding rule and transmit any sequence of \black{their} choice. This model is referred to as a \emph{Byzantine multiple access channel (BMAC)}. \black{The deviating users are controlled by a common adversary, and their transmissions may therefore be correlated.}
 The set of users the adversary controls may be any one of the sets \black{in the adversary structure} $\cA\subseteq 2^{[1:k]}$, where $2^{[1:k]}$ denotes the power set of $[1:k]$. The other users and the decoder are unaware of the identity of the set $\cQ$ of users, $\cQ\in \cA$, that the adversary controls.  The adversarial \black{users do} not know the input of the other \black{(honest) users} and \black{ they remain} fixed during the entire transmission block (cf. Section \ref{Akac}). In the sequel, we refer to the users in this set $\cQ\in \cA$ which the adversary controls as the adversarial users and the other users as trustworthy. \vspace{0.4cm}

    \begin{figure*}
    \centering
    \subfigure[]{
    \begin{tikzpicture}[scale=0.5]
    	\draw[] (4.1+4,4-6) rectangle ++(1.3,1.5) node[pos=.5]{$W$};
    	\draw[->] (3.4+4,5.25-6) -- ++(0.7,0);
    	\draw[red,->] (3.4+4,4.25-6) -- ++(0.7,0);
    	\draw[red,-] (3.4+4,4.25-6) -- ++ (0,-1);
    	\draw[-] (3.4+4,5.25-6) -- ++ (0,1);
    	\draw[red,-] (2.4+4, 3.25-6) node[left]{\textcolor{red}{${y}$}} --  ++(1,0) ;
    	\draw[-] (2.4+4,6.25-6) node[anchor=east]{$f_{\one}({m}_{\one})$} -- ++ (1,0);
    	\draw[->] (5.4+4,4.75-6) --  ++ (0.5,0)node[anchor= west] {$z$};
    \end{tikzpicture}
    \label{fig:dec2a} 
    }
    \subfigure[]{
    \begin{tikzpicture}[scale=0.5]
    	\draw[] (4.1-4,4-6) rectangle ++(1.3,1.5) node[pos=.5]{$W$};
    	\draw[red,->] (3.4-4,5.25-6) -- ++(0.7,0);
    	\draw[->] (3.4-4,4.25-6) -- ++(0.7,0);
    	\draw[-] (3.4-4,4.25-6) -- ++ (0,-1);
    	\draw[red,-] (3.4-4,5.25-6) -- ++ (0,1);
    	\draw[-] (2.4-4, 3.25-6) node[left]{$f_{\two}(\tilde{m}_{\two})$} --  ++(1,0) ;
    	\draw[red, -] (2.4-4,6.25-6) node[anchor=east]{\textcolor{red}{$x$}} -- ++ (1,0);
    	\draw[->] (5.4-4,4.75-6) --  ++ (0.5,0)node[anchor= west] {$z$};
    \end{tikzpicture}
    \label{fig:dec2bb} 
    }
    \subfigure[]{
    \begin{tikzpicture}[scale=0.5]
    	\draw[] (4.1+4,4-6) rectangle ++(1.3,1.5) node[pos=.5]{$W$};
    	\draw[->] (3.4+4,5.25-6) -- ++(0.7,0);
    	\draw[red,->] (3.4+4,4.25-6) -- ++(0.7,0);
    	\draw[red,-] (3.4+4,4.25-6) -- ++ (0,-1);
    	\draw[-] (3.4+4,5.25-6) -- ++ (0,1);
    	\draw[red,-] (2.4+4, 3.25-6) node[left]{\textcolor{red}{$\tilde{y}$}} --  ++(1,0) ;
    	\draw[-] (2.4+4,6.25-6) node[anchor=east]{$f_{\one}(\tilde{m}_{\one})$} -- ++ (1,0);
    	\draw[->] (5.4+4,4.75-6) --  ++ (0.5,0)node[anchor= west] {$z$};
    \end{tikzpicture}
    \label{fig:dec2c} 
    }
    \subfigure[]{
    \begin{tikzpicture}[scale=0.5]
    \draw[teal] (3, 8.5) node{\fbox{\textcolor{black}{$\red{\tilde{X}\tilde{Y}}-\red{Y}-XZ$}}};
    	\draw (4.1,4) rectangle ++(1.3,1.5) node[pos=.5]{$W$};
    	\draw[->] (3.4,5.25) -- ++(0.7,0);
    	\draw[red,->] (3.4,4.25) -- ++(0.7,0);
    	\draw[red,-] (3.4,4.25) -- ++ (0,-1);      
    	\draw[-] (3.4,5.25) -- ++ (0,1);
    	\draw[red,-] (2.4, 3.25)  --  ++(1,0) node[right]{$Y \,\,\textcolor{orange}{\boxed{y}} $};
    
    	\draw[red] (1.4, 3.25-0.6) rectangle ++(1,1.2);
    	\draw[red,->] (1.4-0.7, 3.25-0.4) node[anchor=east]{ $\tilde{Y}$} -- ++ (0.7,0);	
    
    	\draw[orange] (-0.4, 3.25+0.7) node[anchor=east]{ $\textcolor{orange}{\boxed{ f_{\one}(\tilde{m}_{\one})}}$};
    	\draw[orange] (-0.4, 3.25-0.7) node[anchor=east]{ $\textcolor{orange}{\boxed{ f_{\two}(\tilde{m}_{\two})}}$};
    
    	\draw[red,->] (1.4-0.7, 3.25+0.4) node[anchor=east]{ {$\tilde{X}$}} -- ++ (0.7,0);
    	\draw[-] (2.4,6.25) node[anchor=east]{$\textcolor{gray}{\boxed{{f_{\one}(m_{\one})}}}\,\, X$} -- ++ (1,0);
    	\draw[->] (5.4,4.75) --  ++ (0.5,0)node[anchor= west] {$Z\,\,\textcolor{gray}{\boxed{z}}$};
    \end{tikzpicture}
    \label{fig:dec2b} 
    }
    \caption{The channel output $z$ is such that the tuples $(f_{\one}(m_{\one}), y, z)$, $(x, f_{\two}(\tilde{m}_{\two}), z)$ and $(f_{\one}(\tilde{m}_{\one}), \tilde{y}, z)$ are consistent according to channel law for some $y$, $x$ and $\tilde{y}$ as shown in (a), (b) and (c) respectively. Suppose 
    	$(f_{\one}(m_{\one}), {y}, z, {f_{\one}(\tilde{m}_{\one})}, {f_{\two}(\tilde{m}_{\two})})\in T^n_{X{Y}Z{\tilde{X}}{\tilde{Y}}}$ is such that $I({\tilde{X}}{\tilde{Y}};XZ|{Y})$ is small ({\em i.e.} the Markov chain $\tilde{X}\tilde{Y}-{Y}-XZ$ approximately holds). Then, subfigure (d) is a plausible explanation where user \one is trustworthy with input message $\mo$, and  $\tilde{m}_{\one}$ and $\tilde{m}_{\two}$ can be explained by an attack strategy of user \two (compare subfigures (d) and (a)).} \label{fig:dec2}
    \end{figure*}
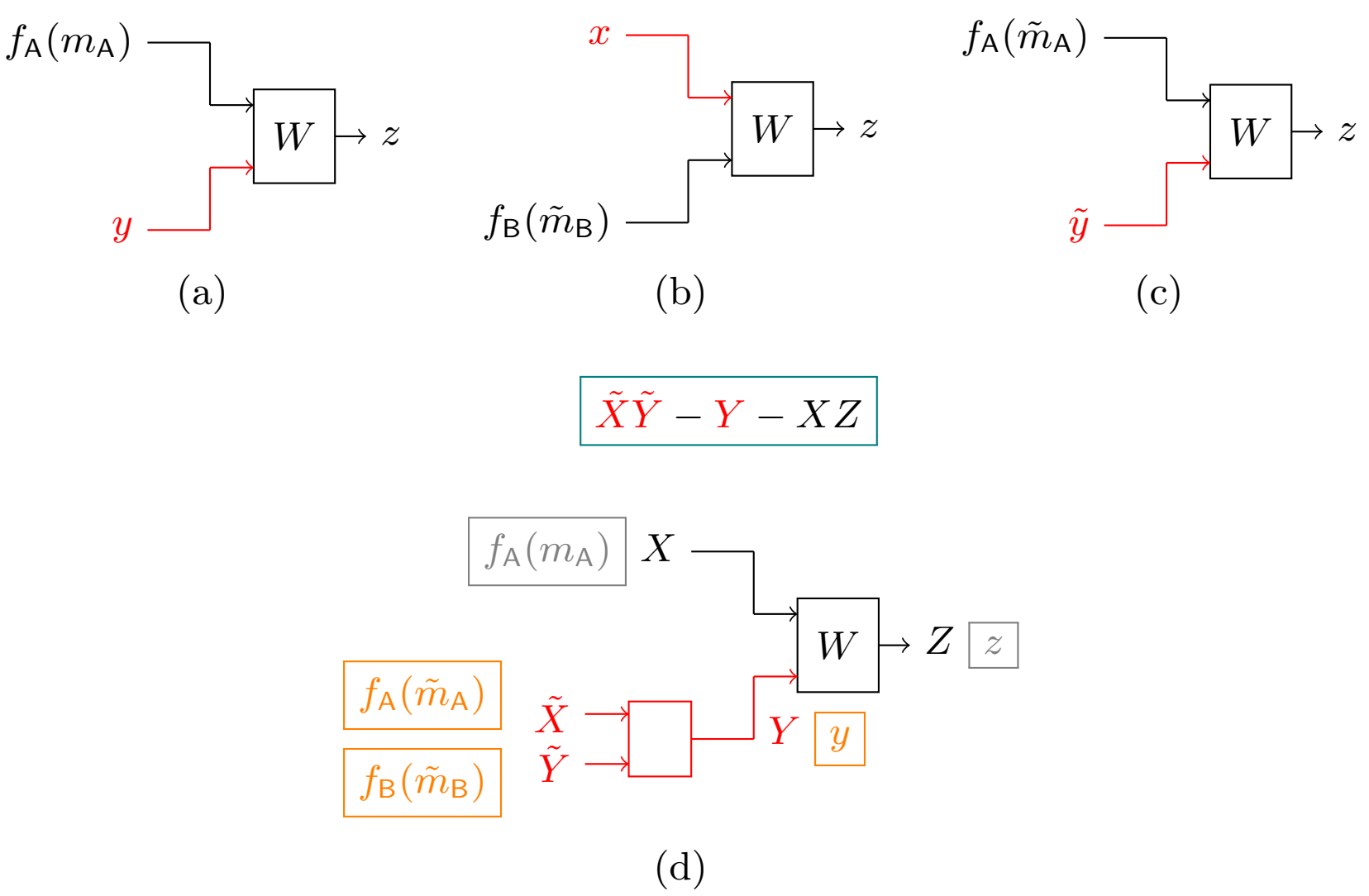

\begin{table}
\caption{\textcolor{black}{Decoding guarantees for different models of Byzantine MACs}}
\label{codestable}
\begin{tabular}{p{1.8cm}|p{3.5cm}|p{3.5cm}|p{3.5cm}}
  \toprule
 & {\bf{Reliable Communication}}&
   {\bf{Authenticated Communication}}&
   {\bf{Communication with Adversary Identification}}\\
  \midrule
Decoding Requirements&
\textcolor{black}{Correctly decode all users' messages when all users are trustworthy. In the presence of adversarial users, correctly decode the messages of all trustworthy users.}&
\textcolor{black}{Correctly decode all users' messages when all users are trustworthy. Otherwise, either correctly decode the trustworthy users' messages or declare an error, without identifying the adversarial user.}&
\textcolor{black}{Correctly decode both users' messages when both users are trustworthy. If one user is adversarial, either correctly decode the trustworthy user's message or identify the adversarial user (in a two-user Byzantine MAC).}
\\
  \bottomrule
\end{tabular}
\end{table}\vspace{0.3cm}

 In the next sections, we review the main capacity results for classical Byzantine multiple access channels.
    Several different notions 
of decoding objective have been considered in the Byzantine communication models, \black{namely, } reliable communication, authenticated communication, and communication with adversary identification.
As Table \ref{codestable} shows, each of these three
\black{models have different decoding requirements and} address different aspects of communication in the presence of adversaries, providing varying levels of security and reliability.  (A detailed comparison of these models is given in the journal paper \cite{Sa/Ba/De/Pr2}).

\Cref{rc} reviews   key results on reliable message transmission capacity and code concepts
 proposed for achieving reliable communication.
\Cref{aut} consists
of a  review of 
\black{authenticated communication in} classical Byzantine MAC.
\Cref{iden} treats the problem of communication with adversary identification.


\section{Reliable Communication}
\label{rc}

We begin with the original communication model  treated by
\cite{Sa/Ba/De/Pr} and \cite{Sa/Ba/De/Pr2}.

\subsection{Randomized Codes and
Deterministic Codes}\label{randet}


\begin{definition}[Randomized code]\label{defn:random_code_k}
An $(\nummsg_1,\nummsg_2,\ldots, \nummsg_k,n)$ randomized {\em code} for the \bmac $(W,\cA)$ consists of the following: 
\begin{enumerate}[label=(\roman*)]
\item $k$ message sets, $\mathcal{M}_i = \{1,\ldots,\nummsg_i\}$, $i=1,2,\ldots,k$,
\item $k$ independent randomized encoders, $F_{i}:\mathcal{M}_i\rightarrow \mathcal{X}_i^n$,  where $F_i\sim P_{F_i}$ takes values in {$\mathcal{F}_i\subseteq\{g:\cM_i\rightarrow \cX_i^n\},\, i =1,2,\ldots,k$} and
\item a decoder, $\phi:\mathcal{Y}^n\times\mathcal{F}_1\times\mathcal{F}_2\times\ldots\times\mathcal{F}_k\rightarrow\mathcal{M}_1\times\mathcal{M}_2\times\ldots\times\mathcal{M}_k$ where \\$\phi(y,F_1,F_2,\ldots,F_k)$ $ = $ $ (\phi_1(y,F_1,F_2,\ldots,F_k),$ $ \phi_2(y,F_1,F_2,\ldots,F_k),$ $ \ldots,$ $ \phi_k(y,F_1,F_2,$ $ \ldots,F_k))$ for some deterministic functions $\phi_i:\mathcal{Y}^n\times\mathcal{F}_1\times\mathcal{F}_2\times\ldots\times\mathcal{F}_k\rightarrow \mathcal{M}_i, \, i = 1,2,\ldots,k$.
\end{enumerate}
\end{definition}

Let $P_{e, \cQ}$ denote the average probability of error when the adversary controls the set $\cQ\in \cA$  of users.
\begin{align}\label{eq:avg_error_k_users}
P_{e, \cQ} = \max_{x_{\cQ}}\frac{1}{(\prod_{i\in \cQ^c}N_i)}\sum_{m_{\cQ^c}\in \cM_{\cQ^c}}P\inp{\inb{\phi(Y)_{\cQ^c} \neq m_{\cQ^c}}|X_{\cQ^c} = f_{\cQ^c}(m_{\cQ^c}), X_{\cQ} = x_{\cQ}}.
\end{align}
The average probability of error $P_{e}$ is given by  
\begin{align*}
P_{e} =\max_{\cQ\in \cA}P_{e, \cQ}.
\end{align*}

\begin{definition}[(Strong) Deterministic code]\label{code:det_kuser}
An $(\nummsg_1,\nummsg_2,\ldots, \nummsg_k,n)$ deterministic {\em code} for the \bmac $(W, \cA)$ consists of: \vspace{-0.25em}
\begin{enumerate}[label=(\roman*)]
\item $k$ message sets, $\mathcal{M}_i = \{1,\ldots,\nummsg_i\}$, $i\in \{1,2,\ldots, k\}$,
\item $k$ encoders, $f_{i}:\mathcal{M}_i\rightarrow \mathcal{X}_i^n$, $i\in \{1,2,\ldots, k\}$, and
\item a decoder, $\phi:\mathcal{Y}^n\rightarrow \mathcal{M}_i.$ 
\end{enumerate}
\end{definition}

The average probability of error $P_{e}$ is given by  
\begin{align*}
P_{e} =\max_{\cQ\in \cA}P_{e, \cQ}.
\end{align*}

We  emphasize that the decoder is unaware of the identity of the
set $Q \in A$ of users controlled by the adversary


  A defining feature of the Byzantine setting is that the decoder must operate reliably under multiple scenarios:
  Consider, for instance, a 2-user Byzantine MAC.
  We distinguish between the event when both users are trustworthy, the event when user 1 is adversarial, and the event when user 2 is adversarial.
    
    
    \medskip
    To illustrate the fundamental difficulty of this problem, \cref{fig:dec2} shows several possible explanations of a received sequence $z$. The same output may be consistent with multiple combinations of transmitted sequences, including scenarios involving adversarial behavior. This ambiguity lies at the heart of adversary identification.
    This observation highlights that the decoder must not only recover messages, but also resolve competing hypotheses about adversarial behavior.

 \begin{definition}[Random code and Deterministic code Capacity region {\cite{Sa/Ba/De/Pr,Sa/Ba/De/Pr2}}]
    A random or deterministic, respectively, rate pair $(R_1,R_2,\cdots,R_k)$ is achievable if there exists a sequence of random or deterministic, codes, respectively, such that $P_e \to 0$ as $n \to \infty$. The randomized coding capacity  region ${\cC}_{\deterministic}$ or the  deterministic coding capacity region ${\cC}_{det}$, respectively, is the closure of all achievable random or deterministic rate pairs, respectively.
    \end{definition}


    
    

    

 \subsection{Symmetrizable Byzantine-MAC}   
\label{SYMM}

\begin{definition}
For a non-empty set $T\subseteq [1:k]$, we say that a \bmac $(W, \cA)$  is {\em\textbf{$T$-symmetrizable}}  if there exist sets $\cQ, \cQ'\in  \cA$, not necessarily distinct, satisfying $\cQ\cap T = \cQ'\cap T = \emptyset$,  and  a pair of conditional distributions $P_{X_{\cQ}|X_{T\cup (\cQ\setminus \cQ')}}$ and $P'_{X_{\cQ'}|X_{T\cup (\cQ'\setminus \cQ)}}$ such that
\begin{align}
&\sum_{x'_{\cQ}\in \cX_{\cQ}}P_{X_{\cQ}|X_{T\cup (\cQ\setminus \cQ')}}(x'_{\cQ}|x_{T}, x_{\cQ\setminus \cQ'})W(y|x'_{\cQ}, \tilde{x}_{T}, x_{\cQ'\setminus \cQ}, x_{(T\cup \cQ \cup \cQ')^c})\nonumber\\
 &= \sum_{\tilde{x}_{\cQ'}\in \cX_{\cQ'}}P'_{X_{\cQ'}|X_{T\cup (\cQ'\setminus \cQ)}}(\tilde{x}_{\cQ'}|\tilde{x}_{T}, x_{\cQ'\setminus \cQ})W(y|\tilde{x}_{\cQ'}, {x}_{T}, x_{\cQ\setminus \cQ'}, x_{(T\cup \cQ \cup \cQ')^c})\label{eq:thirty}
\end{align}
for all $x_{T}, \tilde{x}_{T}\in \cX_{T}$, $x_{\cQ\setminus \cQ'}\in \cX_{\cQ\setminus \cQ'}$, $x_{\cQ'\setminus \cQ}\in \cX_{\cQ'\setminus \cQ}$, $x_{(T\cup \cQ \cup \cQ')^c}\in \cX_{(T\cup \cQ \cup \cQ')^c}$ and $y\in \cY$.
We say that a \bmac $(W, \cA)$ is {\em \textbf{symmetrizable}} if it is $T$-symmetrizable for any $T\neq\emptyset$.
\end{definition}

\begin{theorem}
    \label{achievability_k_users_final}
For a $k$ user \bmac $(W, \cA)$, the interior of the deterministic coding capacity region is empty if and only if it is symmetrizable. Furthermore, when $(W, \cA)$ is not symmetrizable,  $\mathcal{C}_{\deterministic} = \mathcal{C}_{\random}$.
\end{theorem}

\subsection{Capacity Region}

\begin{theorem}

Let $\mathcal{R}$ be the closure of the set of all random rate tuples $(R_1,R_2,\ldots,$ $ R_k)$ such that for some $p(u)p(x_1|u)p(x_2|u)\ldots p(x_k|u)$, the following conditions hold for all $\cQ\in\cA$ and $\cJ\subseteq \cQ^c$,
\begin{align}\label{rate_region_k_user}
\sum_{j\in \cJ}R_j\leq \min_{q(x_{\cQ}|u)}I\inp{X_{\cJ};Y|X_{(\cQ\cup\cJ)^c}, \pink{U}}
\end{align}
where the mutual information above is evaluated using the joint distribution
$p(u) q(x_{\cQ}|u)\prod_{j\in\cQ^c}p(x_j|u)W(y|x_{\cQ},x_{\cQ^c})$. 
\end{theorem}

To show the 
achievement, random encoder technique
is used for the channel input;
the decoding set $D_{m_1,m_2,\ldots, m_k}  \subseteq \mathcal{Y}^n$ of the message tuple $\inp{m_1,m_2,\ldots, m_k}\in \cM_1\times \cM_2\times\ldots\times \cM_k$ is defined as the intersection of the sets ${m_i}, , i\in [1:k]$, {\em i.e.}, $D_{m_1,m_2,\ldots, m_k}:= \cap_{i=1}^{k}D^{i}_{m_i}$, where the sets $D^{i}_{m_i},, i\in [1:k]$
is built with similar approach like in \cite{Cs/Na} for AVC and in \cite{Ahl/Ca} for MAC with known adversary.

Recall that in  \cite{Cs/Na}, the decoding
set  $D_m$ is built such way: at first define a set such that
the channel output, when $m$ has been sent, will be included
in this set with certainty for all adversarial input. Then, eliminating
ambiguous sequences, i.e.,  eliminating all the sequences in this
set when the decoder cannot disambiguate if
a $m'\not= m$ has been sent.
In
\cite{Ahl/Ca}, an 2-user AV-MAC decoding set $D_{m_1,m_2}$ is \black{also built in a similar manner}: at first define a set such that
the channel output, when $(m_1,m_2)$ has been sent, will be included
in this set with certainty.
Next, eliminating
ambiguous sequences,  when the decoder cannot disambiguate if
a ${m_1}'\not= m_1$ or a ${m_2}'\not= m_2$ has been sent.


The construction of decoding set $D_{m_i}$ for a 3-user non-symmetrizable
Byzantine-MAC \black{follows} the same strategy:
\black{Consider} the decoding set $D_{m_3}$ for sender 3's message.
At first, build a set which
contains all candidates for outputs of $m_3$,
i.e., when this sequence can be ``explained" by
the case when sender 3 sends $m_3$, and one of
the other users, say,
sender 1,
for instance, is the adversarial one, and sender 
2 sends some message $m_2$.
Next eliminating
ambiguous sequences:\\
1) eliminating all the sequences in this
set that  can be also explained by
the case when 
sender 3 sends a message ${m_3}'\not= m_3$ while
sender 
2 sends $m_2$.\\
2) eliminating all the sequences in this
set that  can be explained by
the case when 
sender 3 sends a message ${m_3}'\not= m_3$ while
sender 
2 sends some message ${m_2}'\not= m_2$.\\
3) eliminating all the sequences in this
set that   can be explained by
the case when 
sender 3 sends a message ${m_3}'\not= m_3$ while
sender 
2 is actually the adversarial one, while  sender 1
sends some message $m_1$.

This coding concept can be extended to $n$-user non-symmetrizable
Byzantine-MAC:
To construct a decoding set $D_{m_1}$ for sender 1's message,
build a set which
contains all candidates that can be explained by
the case when sender 1 sends $m_1$, and one of
the other users, say,
sender 2,
for instance, is the adversarial one.
Next, eliminate all sequences that can be
explained by
an alternative input of sender 1
when  sender 2 is, in fact,
the adversarial one.
Next, eliminate all sequences 
that can be
explained by
an alternative input of sender 1 
when another sender is actually 
the adversarial one.

\section{Authentication}
\label{aut}
A variation of this problem was considered in \cite{NehaBDP19}
and \cite{NehaBDP21}.
In many practical systems, one wishes to combine  two objectives: achieve high‐rate reliable transmission and ensure that an adversarial user cannot corrupt an trustworthy user's message without being detected. This gives rise to the notion of authenticated communication, which blends the classical requirement of reliable decoding (message transmission) with the security requirement of authentication. 
An authentication scheme does not necessarily increase the rate of information transfer; rather, it augments any underlying transmission protocol with the capability to detect tampering. Concretely, an authenticated receiver must satisfy two requirements:
 \begin{itemize}
  
    \item  No False Rejects: If the sender behaves honestly and the channel is in its ``no‐attack'' mode, the receiver accepts and outputs the correct message.
    
  \item  No Undetected Forgeries: Under any adversarial manipulation - no matter how powerful - the receiver either rejects (aborts) or outputs the truly transmitted message; it must never be fooled into accepting an incorrect message without detecting the attack.
  \end{itemize}

The 
problem of authentication
bases on the observation that it is better to know
what you do not know than it is to be wrong. We assume that
an adversary may or may not be present; if it is not present,
then the receiver should be able to decode the message at
the maximum possible rate; if it is present, then the receiver
should be able to detect the adversary, even if it cannot decode
the message. Thus, when a message is decoded, the receiver
can be certain that this message is authentic. Moreover, this
must be accomplished without any pre-shared key.

\begin{figure}
        \centering
        \includegraphics[width=0.80\linewidth]{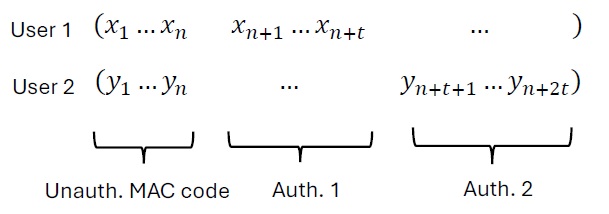}
        \caption{\label{authcode}3-phase scheme for authenticated communication over MAC. The
tag blocklengths t of the authentication phases are asymptotically smaller
than the blocklength n of the preceding unauthenticated communication phase.}
    \end{figure}

The code concept \black{is based} on a 3-phase scheme (cf. Fig. \ref{authcode}). In the first phase  a standard MAC code is used to
communicate the messages, albeit without any guarantee that
the decoded messages are indeed what the senders intended.
In the second phase, user 1 authenticates its message to
the decoder while treating user 2 as a potential adversary;
simultaneously, user 2 is required to send a certain
``no-attack''
symbol from its channel input alphabet throughout this phase
with deviation from this amounting to adversarial behavior.
In phase 3, the roles of the users are reversed,
 user 2 authenticates its message to
the decoder while treating user 1 as a potential adversary.

\subsection{Problem Formulations}

\textbf{Channel model:}
 A two-user MAC  
  \begin{itemize}
    \item 
    Honest case: Both users behave normally; the receiver must decode both messages reliably.
   \item Adversarial case: One user may be controlled by an adversary. The receiver must either
        decode the trustworthy user's message without error, or
      detect the attack and output an ``abort'' symbol.
    \end{itemize}        
  
\textbf{Requirements}
  \begin{itemize}
    \item     
Security: When a trustworthy encoder intends send $m\in M$, and the decoder receive a $m'$, the decoder must never accept if
$m'\not= m$.
  \item  Reliability: If no attack, the decoder must accept with high probability $m'=m$.
  \end{itemize} 
  
 \textbf{Goal:} 
    Characterize the region of message‐rate pairs
    $(R_1,R_2)$ that can be simultaneously achieved under both scenarios.

\subsection{System Model and Authenticated Communication Codes}
\label{OVERW}

Consider the two-user MAC . The memoryless channel  $W_{Z|XY}$  has input alphabets
X, Y, and output alphabet Z. We consider stochastic
encoders which use independent private randomness.

\begin{definition}[authenticated communication code {\cite{NehaBDP19}
\cite{NehaBDP21}}]
    An $(N_1,N_2,n)$ (deterministic) authenticated communication for the
    multiple access channel $W_{Z|XY}$  consists of:
    \begin{enumerate}[label=(\roman*)]
    \item Message sets $\cM_i=\{1,\dots,N_i\}$ for $i=1,2$,
    \item Encoders $f_1:\cM_1\to \cX^n$ and $f_2:\cM_2\to \cY^n$,
    \item A decoder $\phi:\cZ^n\to (\cM_1\times \cM_2)\cup\{\perp,\perp\}$.
    \end{enumerate}
    \end{definition}

\subsection{Overwritability}

We consider a 2-users model.
 We say that user 1 can
overwrite user 2 if there is a distribution 
$P_{X'|XY}$
such that
\[\sum_{x'\in X} P_{X'|XY}(x'|xy)W(z|x',y') = 
W(z|x,y)  \]
 for all $x$, $y$, $y'$, $z$. In other words, user 1 can
overwrite user 2 if it (user 1) can replace any symbol $y'$ sent
by user 2 with any symbol $y$ of its choosing while pretending
to send any symbol $x$, again, of its choosing. In doing this,
user 1 does not have knowledge of the symbol  $y'$ transmitted
by user 2. Overwritability of user 1 by user 2 can be defined
in a similar manner. The overwritable condition is an extension of  symmetrizability
 (cf. Section \ref{SyandSp}).

\subsection{Capacity results}

The main results are stated below:
\begin{theorem}\label{autcap}
Let $C_{non-adv}$ denote the non-adversarial capacity region of
the DM-MAC with stochastic encoders under the maximum
probability of error criterion.

  \begin{enumerate}[label=(\roman*)]
    \item If the channel is such that neither user can
overwrite the other, then $C_{auth} = C_{non-adv}$;
if it is overwritable, then $C_{auth} = 0$.
\item If user 1 can overwrite user 2, but user 2 cannot overwrite
user 1, then $C_{auth} = \{(R, 0)|R \in  C_{non-adv}\}$. 
\item If
user 2 can overwrite user 1, but user 1 cannot overwrite user
2,  then $C_{auth} = \{(0,R)|R \in  C_{non-adv}\}$. 
\item When both
users can overwrite each other, then $C_{auth} = 0$.
\end{enumerate}
\end{theorem}

The coding
scheme consists of three phases: a DM-MAC code
(of blocklength n, say) with stochastic encoders for non-adversarial
users with maximum probability of error vanishing
as $n\rightarrow \infty$.
\black{As a first step, the achievability proof shows}  that for any non-overwritable
channel, the authentication capacity is positive.
Then it shows that when the encoder and decoder
have access to a small amount of shared randomness,
unknown to the adversary,
they can achieve the no-adversary capacity.
This means that any positive rate code can be used to establish shared
randomness, and thereby achieve the no-adversary capacity.
The techniques for this part are similar to
the weak deterministic coding technique
for non-symmetrizable
 classical AVCs given in \cite{Ahl1}.
This phase
 is followed by two authentication phases,
one for each user to detect if their are trustworthy or not.
The authentication
involves transmitting an identification
code for the identity channel (cf. \cite{Ah/Du}) using
an authenticated communication code. Using the observation
that given a deterministic authenticated communication code
from \cite{Ko/Kl} that has a small average probability of error, i.e., similar to the identification
problem  \cite{Ah/Du}, error probability scales vanish
doubly exponentially, we can
construct a stochastic authenticated communication code with
a slightly lower rate and with a small maximal probability
of error. 

\subsection{Discussion}
\label{autdis}
The authentication problem studied here borrows
features from the identification setup over non-AVCs \cite{Ah/Du} as
well as that over AVCs \cite{Bo/De}. On the one hand, as also in
the identification setup over the non-AVC, we
desire that the true message be accepted under no attack.
On the other hand, as in the identification setup over the AVC, we desire that a wrong candidate message be rejected
regardless of the channel state vector. However, unlike the
identification setup, when the system is under attack, authentication does not
have any requirement on the decoder behavior if the candidate
message is the same as the true message.

Theorem \ref{autcap} gives us insight into the nature
of the result of \cite{Ko/Kl}.
The result of \cite{Ko/Kl} is that whenever the DM-AVC is not overwritable,
the authenticated communication capacity can be achieved, and it is zero otherwise. Theorem \ref{autcap} suggests  that an alternative
architecture (but not an alternative proof) to achieve this
result is to first perform unauthenticated communication using
a good code for the nominal channel and follow this up
with a short authentication phase which does not affect the
overall rate. Arguably, this explains the nature of the result
of  \cite{Ko/Kl} - whenever authenticated communication is possible
at all, since authentication can be performed for a set of
doubly-exponentially growing number of messages, the rate
of communication is simply governed by the largest rate of
unauthenticated communication possible, i.e., the Shannon
capacity of the nominal channel.

\section{Communication With Adversary Identification}
\label{iden}

\black{We consider the following question in this section}:
Is it possible to design a coding and decoding scheme for a two-user multiple access channel (MAC) so that, whenever one user acts maliciously, the decoder either
  \begin{itemize}
    \item   
    correctly identifies the misbehaving user, 
    \item or
    still recovers the trustworthy user’s message reliably,
    while when both users behave honestly it decodes both messages with high probability?
\end{itemize}

We call this formulation ``communication with adversary identification'' over a byzantine-MAC. Concretely:
 \begin{itemize}
    \item  
    The channel is used by two encoders, each choosing a message and sending a codeword.
    One of the users may be permanently controlled by an 
    \item adversary who can arbitrarily deviate from the prescribed encoding rule.
\item    The receiver must output either:
 \begin{itemize}
    \item  
    a pair of messages (one from each user), 
    \item or
        the identity of the deviating user.
\end{itemize} \end{itemize}

The decoding requirements are:
 \begin{itemize}
    \item   
    If both users are trustworthy, the decoder outputs the correct message pair w.h.p.
   \item If exactly one user is adversarial, then w.h.p. the decoder either identifies that user as the deviator, or else correctly recovers the trustworthy user’s message.
  \item  No constraint is imposed when both users deviate.
\end{itemize}

\subsection{System Model and Adversary Identifying Codes}
\label{OVERW2}

Consider the two-user MAC . The memoryless channel  $W_{Z|XY}$  has input alphabets
X, Y, and output alphabet Z. We consider stochastic
encoders which use independent private randomness.

    \begin{definition}[Adversary identifying code {\cite{Sa/Ba/De/Pr,Sa/Ba/De/Pr2}}]
    An $(N_A,N_B,n)$ (deterministic) adversary{-}identifying code for the
    multiple access channel $W_{Z|XY}$  consists of:
    \begin{enumerate}[label=(\roman*)]
    \item Message sets $\cM_i=\{1,\dots,N_i\}$ for $i=A,B$,
    \item Encoders $f_A:\cM_A\to \cX^n$ and $f_B:\cM_B\to \cY^n$,
    \item A decoder $\phi:\cZ^n\to (\cM_A\times \cM_A)\cup\{A,B\}$.
    \end{enumerate}
    \end{definition}

Here $\{A,B\} $ are additional symbols so that the decoder outputs either a pair of messages or declares one of the users as adversarial.

\subsection{Relation to Prior Models}

Our adversary-identification model lies strictly between 
the extreme models of reliable communication and authenticated communication in a Byzantine-MAC:

    It relaxes the requirement of always correctly decoding both users (as in reliable communication), allowing the system to focus on preserving the trustworthy user’s data or catching the adversary.
    It also strengthens authenticated communication by demanding identification of the tampering user, not merely a detection of interference.
    
Decoding strategies for reliable or authenticated MACs typically handle one user at a time
- either by viewing the other user's input as an AVC state or by appending short authentication phases. In contrast, identifying which user deviates requires jointly reasoning about both users' codewords and the channel noise. This simultaneous treatment leads to fundamentally new design and analysis challenges.

 \subsection{Spoofable Byzantine-MAC}   
\label{SPOOF}

We consider a 2-users Byzantine-MAC $W_{Z|XY}$. 
We say that $W_{Z|XY}$ is A-spoofable if there exist conditional distributions ${Q_{Y|\tilde{X}\tilde{Y}}}$ and ${Q_{X|\tilde{X}X'}}$ such that $\forall\,x', \tilde{x}\in \cX,\, \tilde{y}\in \cY,\, z\in \cZ,$
\begin{align}\label{eq:spoof1}
&\sum_{y}Q_{Y|\tilde{X}\tilde{Y}}(y|\tilde{x},\tilde{y})W(z|x',y) \nonumber\\
&= \sum_{y}Q_{Y|\tilde{X}\tilde{Y}}(y|x',\tilde{y})W(z|\tilde{x},y) \nonumber\\
& = \sum_{x}Q_{X|\tilde{X}X'}(x|\tilde{x},x')W(z|x,{\tilde{y}}).
\end{align}

We say that $W_{Z|XY}$ is B-spoofable if there exist conditional distributions
$Q_{Y |\tilde{X}\tilde{Y}}$
and $Q_{Y |\tilde{X}Y'}$ such that $\forall$
$\tilde{x}\in X$, $y'$, $\tilde{y}\in Y$, 
$z \in Z$
\begin{align*}
& \sum_{x}Q_{X |\tilde{X}\tilde{Y}}(x|\tilde{x},\tilde{y})
W_{Z|XY} (z|x, y')\\
&=\sum_{x}Q_{X |\tilde{X}\tilde{Y}}(x| \tilde{x},y')
W_{Z|XY} (z|x,\tilde{y})\\
&=\sum_{y}Q_{Y |\tilde{Y}Y'}(y|\tilde{y},y')
W_{Z|XY} (z|\tilde{x},y)
\text{ .}
\end{align*}

A Byzantine-MAC is spoofable if it is either A-spoofable or B-spoofable.

 \begin{theorem} If a Byzantine-MAC is A-spoofable (resp. B-spoofable), deterministic communication with adversary identification from user-A
(resp. user-B) is impossible.  If a Byzantine-MAC is neither A-spoofable nor B-spoofable, then its capacity region 
${\cC}_{det}$ has
a non-empty interior, i.e., both users can communicate deterministically reliably with adversary identification at positive rates.
\end{theorem}     

\subsection{Capacity Region}

 \begin{theorem}

For distributions $P_{\one}$ and $P_{\two}$  over $\cX$ and $\cY$ respectively, we define  $\cP(P_{\one}, P_{\two}) \defineqq \{P_{XY\tilde{X}\tilde{Y}Z}: P_{X\tilde{Y}Z}=P_{\one}\times P_{\tilde{Y}}\times W \text{ for some }P_{\tilde{Y}} \text{ and }P_{\tilde{X}YZ} = P_{\tilde{X}}\times P_{\two}\times W \text{ for some }P_{\tilde{X}}\}$.
Let $R_{1}(P_{\one}, P_{\two})$ be the set of rate pairs $(R_{\one}, R_{\two})$ such that
\begin{align}\label{eq:inner_bd_1}
R_{\one}&\leq \min_{P_{XY\tilde{X}\tilde{Y}Z} \in \cP(P_{\one}, P_{\two})} I(X;Z), \text{ and }\nonumber\\
R_{\two}&\leq \min_{P_{XY\tilde{X}\tilde{Y}Z} \in \cP(P_{\one}, P_{\two}):X\independent Y} I(Y;Z|X).
\end{align} 
Similarly, let $R_{2}(P_{\one}, P_{\two})$ be the set of rate pairs given by
\begin{align}\label{eq:inner_bd_2}
R_{\one}&\leq \min_{P_{XY\tilde{X}\tilde{Y}Z} \in \cP(P_{\one}, P_{\two}):X\independent Y} I(X;Z|Y), \text{ and }\nonumber\\
R_{\two}&\leq \min_{P_{XY\tilde{X}\tilde{Y}Z} \in \cP(P_{\one}, P_{\two})} I(Y;Z).
\end{align}

When $\mathsf{int}(\cC)\neq \emptyset$, we can archive the 
deterministic capacity region:
\begin{align*}
\mathsf{conv}(\cup_{P_{\one}, P_{\two}}\inp{R_1(P_{\one}, P_{\two})\cup R_{2}(P_{\one}, P_{\two})})\subseteq \cC.
\end{align*}
\end{theorem}

Like in the achievement proof for 
reliable message transmission in Section
\ref{rc}, 
the decoding sets $D_A$ and $D_B$
for a 
Byzantine-MAC
are built with similar approach like for AVC:

When a Byzantine-MAC is not spoofable, the users can construct  
a weak deterministic code, i.e., a
two-phased code. In the first phase of the code,
a positive-rate, deterministic adversary identifying code  is used to send
random bits as messages by each (trustworthy) sender to the receiver. This establishes the  amount of shared
randomness required for the randomized adversary identification code in the second phase.

$D_A$ and $D_B$ are
 constructed as follows: At first, the decoder builds
the two sets containing candidate messages for user A and B, respectively.
After that, he eliminates all
ambiguous sequences and eliminates every  sequence in the
set which is a potential adversarial attack; i.e.,
if the messages of the other user cannot 
be disambiguated by this sequence.
The decoded output is as follows.

\begin{align*}
\phi(z) = \begin{cases}(m_a,m_b) &\text{ if }D_A\times D_B = \{(m_a,m_b)\},\\ a &\text{ if }|D_A| = 0, \, |D_B| \neq 0,\\b &\text{ if }|D_A| \neq 0, \, |D_B| = 0\text{ and}\\(1,1)&\text{ otherwise.}
\end{cases}
\end{align*}

\section{Byzantine Distributed Function Computation and Authenticated Communication}
\label{sec:distributed_computation}

Another model with  a weaker decoding guarantee was
proposed in \cite{AnilkumarSPNP25} and \cite{AnilkumarSPNPArxiv}.
In the distributed source coding problem, spatially distributed
users observe correlated sources and send  their encoded observations to a central decoder which, using these
messages and any source observation (side-information) of its
own, attempts to reconstruct all the sources or a function of
them.
   Byzantine distributed function computation shifts
    the focus from recovering all individual messages to computing a prescribed
    function of distributed data. 
    The decoder wants to either
compute a desired function  of the sources or be able to
identify an adversarial user.
    This setting is motivated by scenarios in which a
    central decoder is interested only in a task-relevant quantity, while some of the
    participating users may be malicious.

\subsection{Problem Formulation}

    We consider a distributed source model with $k$ users and one decoder. User
    $i\in[k]$ observes a source sequence $X_i^n$, while the decoder observes side
    information $Y^n$. The random variables
    \[
     (X_{1,t},\ldots,X_{k,t},Y_t), \qquad t\in[n],
    \]
    are assumed to be independent and identically distributed according to a known
    joint distribution $P_{X_1,\ldots,X_k,Y}$.
    
    The decoder is interested in reconstructing a function
    \[
     Z_t = f(X_{1,t},\ldots,X_{k,t},Y_t),
    \]
    or equivalently the sequence $Z^n$. In contrast to classical distributed source
    coding, the main limitation is not the communication rate but the lack of trust.
    Indeed, the users are allowed to send their observations uncoded to the decoder,
under the promise that at most $s\leq k$ users may be controlled by an adversary.
    
    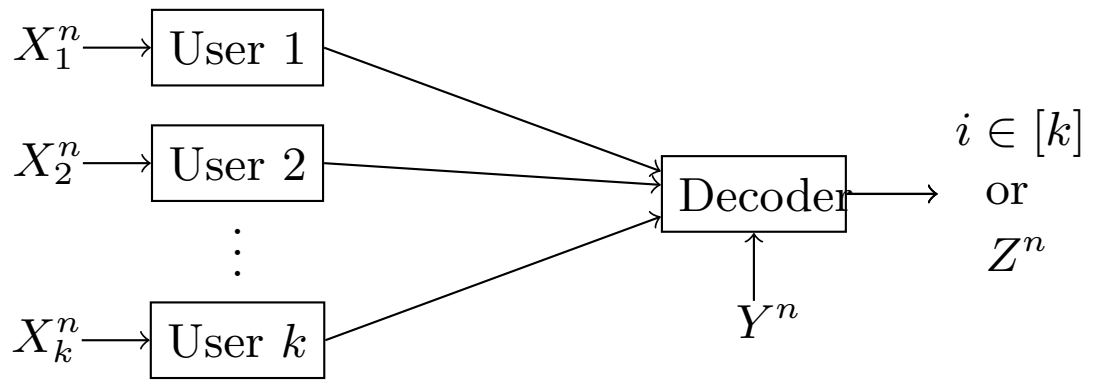
\begin{figure}[tb]
    \centering
    
    \begin{tikzpicture}[node distance=0.25cm, every node/.style={draw, minimum size=0.5cm, align=center}, scale = 0.6]
    
    \node (user1) [rectangle] {\footnotesize User~1};
    \node (user2) [rectangle, below=of user1] {\footnotesize User~2};
    \node (dots) [draw=none, below=of user2, yshift = 0.4cm] {$\vdots$};
    \node (userk) [rectangle, below=of dots, yshift = 0.2cm] {\footnotesize User~$k$};
    
    \draw[->] ($(user1.west)+(-0.75,0)$) -- (user1.west) node[midway, left, draw=none, xshift=-0.1cm] {\footnotesize \(X_1^n\)};
    \draw[->] ($(user2.west)+(-0.75,0)$) -- (user2.west) node[midway, left, draw=none, xshift=-0.1cm] {\footnotesize \(X_2^n\)};
    \draw[->] ($(userk.west)+(-0.75,0)$) -- (userk.west) node[midway, left, draw=none, xshift=-0.1cm] {\footnotesize \(X_k^n\)};
    
    \node (decoder) [rectangle, right=2.8cm of $(user1)!0.5!(userk)$, text width=1cm] {\footnotesize Decoder};
    
    \draw[->] ($(decoder.south)+(0,-0.75)$) -- (decoder.south) node[midway, below, draw=none, xshift = 0.1cm, yshift = -0.1cm] {\footnotesize \(Y^n\)};
    
    \draw[->] (user1.east) -- ($(decoder.west)+(0,0.25)$);
    \draw[->] (user2.east) -- ($(decoder.west)+(0,0.1)$);
    \draw[->] (userk.east) -- ($(decoder.west)+(0,-0.25)$);
    
    \draw[->] (decoder.east) -- ($(decoder.east)+(1,0)$) node[midway, right, draw=none, xshift=0.3cm, yshift=0.4cm] {\footnotesize \(i\in[k]\)};
    \draw[->] (decoder.east) -- ($(decoder.east)+(1,0)$) node[midway, right, draw=none, xshift=0.5cm] {\footnotesize or};
    \draw[->] (decoder.east) -- ($(decoder.east)+(1,0)$) node[midway, right, draw=none,xshift=0.5cm, yshift=-0.4cm] {\footnotesize \(Z^n\)};
    
    \end{tikzpicture}
    \caption{At most $s$ out of the $k$ users are malicious (i.e., controlled by an adversary who has access to the observations of the adversarial users, but no additional side-information). The decoder, with high probability, either identifies an adversarial user or outputs a substantially correct estimate of $Z^n$, where $Z=f(X_1,\ldots,X_k,Y)$. The main result is a characterization of functions $f$ for which this is possible for a given $P_{X_1,\ldots,X_kY}$. If the decoder identifies an adversarial user, that user can be removed to get a new instance of the problem with $k-1$ users of which at most $s-1$ are adversarial and the process repeated.}\label{fig:setup}
    \end{figure}

    If a set $A\subseteq[k]$ of users is adversarial, with $|A|\leq s$, then the
    adversary observes $X_A^n$ and may replace it by an arbitrary reported sequence
    $\bar X_A^n$. The decoder does not know the set $A$ and only knows the upper
    bound $s$ on the number of adversarial users. Thus, the decoder receives
    \[
     \langle \bar X_A^n, X_{A^c}^n\rangle
    \]
    together with its side information $Y^n$.

Here, we use underbar (${\underline{X}}_1,{\underline{X}}_2,{\underline{Y}}$) and $Q_{{\underline{X}}_1{\underline{X}}_2{\underline{Y}}}$ to denote the ``views'' of the decoder (only one of ${\underline{X}}_1,{\underline{X}}_2$ is potentially produced by an adversary, but the decoder does not know which one and the notation reflects this fact). We also use $Q_{{\underline{X}}_1\tilde X_1{\underline{X}}_2\tilde X_2{\underline{Y}}}$ and $\tilde X_1,\tilde X_2$ to denote the underlying true observations which are replaced by the 
adversary. We use overbar $\bar X_1,\bar X_2$ and $W_{\bar X_1|X_1},W_{\bar X_2|X_2}$ to denote the replacement (manipulated) random variables and channels used to produce them, respectively.

\begin{definition} \label{def:s-viable} We say a function $f$ with domain $\cX_1\times\ldots\cX_k\times\cY$ is s-{\em viable} 
if for any collection $\cA_1,\cA_2,\ldots,\cA_m$ of distinct subsets of $[k]$ such that $|\cA_i|\leq s,\, i\in[k]$ and $\bigcap_{i=1}^m \cA_i=\varnothing$, under every joint p.m.f. $Q_{{\underline{X}}_{[k]},{\underline{Y}},\left(\tilde X^{i}_{\cA_i}\right)_{i\in[m]}}$ over $\cX_1\times\ldots\cX_k\times\cY\times\prod_{i=1}^m\prod_{j\in \cA_i}\cX_j$ 
satisfying, for each $i\in[m]$,
\begin{itemize}
\item[(a)] 
$Q_{\ina{\tilde X^i_{\cA_i},\,{\underline{X}}_{{\cA}^c_i}},{\underline{Y}}}=P_{X_{[k]}Y}$, and
\item[(b)] 
${\underline{X}}_{\cA_i} \leftrightarrow \tilde X^i_{\cA_i} \leftrightarrow ({\underline{X}}_{{\cA}^c_i},{\underline{Y}})$,
\end{itemize}
we have, for all $i,i'\in[m]$, 
\[ f\inp{\ina{\tilde X^{i}_{\cA_i},\,{\underline{X}}_{{\cA}^c_i}},{\underline{Y}}} = f\inp{\ina{\tilde X^{i'}_{\cA_{i'}},\,{\underline{X}}_{{\cA}^c_{i'}}},{\underline{Y}}}.\]
\end{definition}

\subsection{Robust Recovery}

    A decoder is a mapping
    \[
     \phi_n:\mathcal X_1^n\times\cdots\times\mathcal X_k^n\times\mathcal Y^n
     \to [k]\cup \mathcal Z^n .
    \]
    Thus, the decoder either outputs an index of a user declared adversarial or an
    estimate $\hat Z^n$ of the desired function sequence.

Let $\gamma>0$. Suppose $\cA\subseteq[k]$ is such that $|\cA|\leq s$ (note that $\cA$ may be empty). When the adversary controls the nodes in $\cA$, based on the observations $X_{\cA}^n:=(X_i^n)_{i\in\cA}$, it produces purported observations $\bar X_{\cA}^n$ which are sent to the decoder. Let $W_{\cA}$ denote the randomized map (not necessarily memoryless) that the adversary uses to map $X_{\cA}^n$ to $\bar X_{\cA}^n$. Both $\cA$ and $W_{\cA}$ are unknown to the decoder (other than the fact that $|\cA|\leq s$). 

To define the error event, let ${\cA}^c=[k]\setminus\cA$ and
\begin{align*}
    \cE_1(\cA) &= (\phi(\langle\bar X_{\cA}^n,X_{{\cA}^c}^n\rangle,Y^n)\in {\cA}^c)\\
    \cE_2(\gamma,\cA) &= \big((\phi(\langle\bar X_{\cA}^n,X_{{\cA}^c}^n\rangle,Y^n)\notin [k])\cap\\
    &\qquad(d_{H}({\phi(\langle\bar X_{\cA}^n,X_{{\cA}^c}^n\rangle,Y^n),Z^n}) > \gamma)\big),
\end{align*}
where the average Hamming distortion $d_H(\hat{z}^n,z^n)$ measures the fraction of locations where the vectors differ.
We define the error event when the adversary controls $\cA$ as $\cE(\gamma,\cA)=\cE_1(\cA)\cup \cE_2(\gamma,\cA).$
Thus, an error occurs if either the decoder identifies a node outside $\cA$ or it outputs an estimate of $Z^n$ which incurs an average Hamming distortion more than $\gamma$. Notice that for $\cA=\varnothing$ (i.e., when the adversary is absent), an error occurs unless the decoder outputs an estimate $\hat{z}^n$ and it is of average Hamming distortion no larger than $\gamma$.
Denote the probability of the error event by
\[ \eta(\gamma,\cA,W_{\cA}) = \mathbb{P}(\cE(\gamma,\cA)),\]
where the probability is evaluated under the joint distribution
\[ P(x_{[k]}^n,y^n,\bar x_{\cA}^n) = \left( \prod_{t=1}^n P_{X_{[k]},Y}(x_{[k],t},y_t)\right) W_{\cA}(\bar x_{\cA}^n|x_{\cA}^n).\]

The error probability of the decoder $\phi$ is defined
as the maximum over the choices available to the adversary
$\epsilon(\gamma, \phi) = \max_{A\subseteq[k]:|A|\leq s}$
$\sup_{W_A}\eta(\gamma,A,W_A)$.

We say that a function f with domain $f(X_1\times X_2 \times\ldots\times X_k\times Y)$ is
s-robustly recoverable if, for all $y > 0$, there is a sequence
of decoders $\phi_n$  such that $\lim_{n\rightarrow\infty}
 \epsilon(\gamma, \phi_n) = 0$.

    The robust recovery requirement has two parts. If no adversary is present, the
    decoder should reconstruct $Z^n$ with vanishing normalized Hamming distortion.
    If an adversary is present, then the decoder should either identify one of the
    adversarial users or still output a sequence $\hat Z^n$ whose distortion from
    $Z^n$ vanishes. In particular, it is considered an error if the decoder identifies
    a trustworthy user as adversarial, or if it outputs a function estimate whose distortion
    is too large.
    
    This criterion captures the task-oriented nature of the model. The decoder is
    not required to reconstruct the observations of all trustworthy users. It only needs to
    recover the function value that is relevant for the task, unless the adversarial
    behavior can be detected.
    \subsection{Main Results}

    The central result of Byzantine distributed function computation is a
    single-letter characterization of the functions that can be robustly recovered (cf. \cite{AnilkumarSPNP25}).    This characterization is expressed in terms of an $s$-viability condition.
    
    Informally, a function is $s$-viable if all adversarial explanations of the same
    decoder observation lead to the same function value. More precisely, suppose
    that different adversarial subsets, each of size at most $s$, could have produced
    the same reported data distribution at the decoder. If the decoder cannot
    distinguish between these explanations, then robust computation is possible only
    if the desired function has the same value under all of them.
    
   \begin{theorem} A function $f$ is $s$-robustly recoverable if and
    only if it is $s$-viable. 
  \end{theorem}  
    
    Hence, the viability condition exactly captures the    information about the sources that remains verifiable in the presence of up to
    $s$ adversarial users.
    
    The converse follows from an indistinguishability argument. If two different
    adversarial scenarios induce the same distribution on the data observed by the
    decoder, then no decoder can reliably distinguish between them. Therefore, the
    function values associated with these scenarios must agree. Otherwise, an
    adversary could force the decoder into producing an erroneous function value
    without being detected.

        The achievability proof is based on type and typicality arguments. At first, consider the  $1$-viable 
    2-users scenario:
We denote by ${\it V}_1$ the set of {\em single-letter} (i.e., $n=1$) joint distributions of the inputs to the decoder that user-1 acting adversarially can induce (using some channel $W_{\bar X_1|X_1}$ to produce its report).
\begin{align*} {\it V}_1=\big\{& Q_{{\underline{X}}_1{\underline{X}}_2{\underline{Y}}}: \exists W_{\bar X_1|X_1} \text{s.t. } 
              Q_{{\underline{X}}_1{\underline{X}}_2{\underline{Y}}}(\underline{x}_1,\underline{x}_2,\underline{y})=
\\&\quad\sum_{x_1} P_{X_1X_2Y}(x_1,{\underline{x}}_2,{\underline{y}})W_{\bar X_1|X_1}({\underline{x}}_1|x_1), \forall {\underline{x}}_1,{\underline{x}}_2,{\underline{y}}\big\}.
\end{align*}

We denote   the empirical distribution (type) of the vector $u^n=(u_1,u_2,\ldots,u_n)$
by ${\bf P}_{u^n}$ and
the total variation distance between p.m.fs $Q$ and $Q'$ by $d_{tv}(Q,Q')$.
For $\delta>0$, let ${\it V}_1^{\delta}=\bigcup_{Q\in{\it V}_1} B(Q,\delta)$, where $B(Q,\delta)=\{ Q': d_{tv}(Q,Q')\leq\delta\}$ is the set of p.m.f.s within total-variation distance $\delta$ of $Q$. Analogously, ${\it V}_2, {\it V}_2^{\delta}$ are defined for user-2.
Let $\delta>0$. For $n\in\mathbb{Z}^+$, the decoder
outputs
\begin{align*}
    \phi_n({\underline{x}}_1^n,{\underline{x}}_2^n,{\underline{y}}^n)=
        \begin{cases}
            1, & {\bf P}\underline{x}_1^n,{\underline{x}}_2^n,\underline{y}^n\in {\it V}_{1}^{\delta}\cap({{\it V}_{2}^{\delta}})^c\\
            2, & {\bf P}{\underline{x}}_1^n,{\underline{x}}_2^n,{{\underline{y}}^n}\in ({{\it V}_{1}^{\delta}})^c\cap{{\it V}_{2}^{\delta}}\\
            g({\underline{x}}_1^n,{\underline{x}}_2^n,{\underline{y}}^n), & \text{otherwise},
        \end{cases}\end{align*}

    This can be interpreted as following: the decoder checks
    whether the empirical distribution of the reported observations and side
    information can be ``explained" by a particular adversarial subset. If exactly one
    such explanation is possible, the decoder identifies a adversarial user. This concept can be extended
 to the $s$-viability scenario.

\subsection{Discussion}
\label{distdis}
  Unlike in reliable communication, where  all messages shall be recovered,
  in the distributed function computation scenario, only the
task-relevant information must be recovered. Consequently, certain functions may
remain robustly computable even when the individual observations of the trustworthy
 users cannot be fully reconstructed.
    
   We consider the role of collusion: For $s=1$, the decoder can often
    use pairwise consistency and side information to verify parts of the users'
    reports. For larger $s$, however, colluding adversaries have more side information
    and can coordinate their reports. The $s$-viability condition captures exactly
    which functions remain invariant under all such indistinguishable manipulations.
    
    Thus, Byzantine distributed function computation provides a natural bridge
    between multi-user information theory, secure distributed computation, and
    task-oriented communication. It shows that robustness against internal
    adversaries depends not only on the channel or source statistics, but also on the
    specific function that the system is designed to compute.
\vspace{0.3cm}






\section{Quantum Byzantine Communication}
\label{sec:quantum_bmac}
In Sections \Cref{rc} through
\ref{sec:distributed_computation}, we have considered
 several  decoding concepts in the Byzantine communication models on classical channels. 
    A natural next step is to extend Byzantine communication models to quantum
    communication systems. While many conceptual ideas from classical Byzantine
    communication remain relevant, the quantum setting \cite{CaiQuantumByzantineMAC} introduces several new
    phenomena that fundamentally alter the interaction between communication,
    decoding, and adversarial behavior. 
    

\subsection{Classical-Quantum Channel Model}
    In the quantum setting, communication takes place over quantum
    channels, where  channel inputs induce quantum output states.
    Unlike classical outputs, quantum states cannot generally be observed without
    disturbance. Consequently, decoding operations themselves become part of the
    adversarial interaction.
    
Quantum channels can transmit both classical and
    quantum information. In this survey, we focus on the transmission of classical
    information over quantum channels, i.e., classical-quantum channels.
    
    For a finite-dimensional complex Hilbert space \( A \), the set of density operators on \( A \) is denoted by  
    $\mathcal{S}(A) := \left\{ \rho \in \mathcal{L}(A) \mid \rho \text{ is Hermitian, } \rho \geq 0, \text{ and } \mathrm{tr}(\rho) = 1 \right\}$,
    Let \( \cX \) be a finite alphabet, and \( A \) be a finite-dimensional complex Hilbert space. The entropy for quantum states $\rho\in \cS(A)$ is the von Neumann entropy, $S(\rho^A)=S(A)_\rho=-\Tr\rho\log\rho$, and by analogy with the Shannon entropy we have also the conditional entropy $S(A|B)_\rho = S(AB)_\rho-S(B)_\rho$ and the mutual information $I(A:B)_\rho = S(A)_\rho+S(B)_\rho-S(AB)_\rho$. A classical-quantum channel is a map \( \mathtt{W}: ´\cX \rightarrow \mathcal{S}(A) \), where \( x \mapsto \mathtt{W}(x) \).
    For a set \(\cX \), a Hilbert space \( A \), a classical-quantum channel \( \mathtt{V}: \cX \rightarrow \mathcal{S}(A) \), and a probability distribution \( P \) on \( \cX \), the quantum mutual information of the channel \( \mathtt{V} \) with input distribution \( P \) is defined as  
    $I(P; \mathtt{V})$ $ :=$ $\sum_{x \in \cX} P(x)$ $ D( \mathtt{V}(x)$ $ \parallel \sum_{x' \in \cX} P(x')$ $ \mathtt{V}(x') )$,
    where $D(\rho \parallel \sigma)$ is the quantum relative entropy between \( \rho \) and \( \sigma \).
    
    In analogy to the classical Byzantine MAC, we now define coding schemes for
    classical-quantum Byzantine multiple access channels. However, unlike in the
    classical setting, decoding is performed through POVM measurements acting on
    quantum states. 
    
  \begin{definition}
A random code for a \( k \)-user Byzantine multiple access classical-quantum channel \( \mathtt{W} \) consists of an encoder  
\[
\Bigl\{ \{ x_i^n(m_i)^{\gamma_i} : \gamma_i \in \Lambda_i \} : m_i \in M_i \Bigr\} \subset X_i^n
\]  
for every sender \( i \in \{1, \dots, k\} \), and a set of positive operator-valued measure (POVM) decoder operators  
\[
\Bigl\{ \left\{ D_{m_i}^{\gamma_i} : m_i \in M_i \right\} : \gamma_i \in \Lambda_i \Bigr\}
\]  
for each sender's message. These are parameterized by a random variable on a finite set \( \Lambda_i \). The interpretation of this definition is that both sender \( i \) and the receiver have access to the outcome of a random experiment uniformly distributed on \( \Lambda_i \).
In contrast to the
    classical setting, such randomness is currently essential for the known coding
    schemes for arbitrarily varying classical-quantum channels and Byzantine
    classical-quantum multiple access channels
    (cf. the discussions in \cref{Keydiff}).

The average probability of decoding error for the message of sender \( i \) is defined as  
\[
1 - \frac{1}{J_n} \frac{1}{|\Lambda_i|} \sum_{m_i \in M_i} \sum_{\gamma_i \in \Lambda_i} \mathrm{tr} \left( {\mathtt{W}}^{\otimes n} \left( x_i^n(m_i)^{\gamma_i} \right) D_{m_i}^{\gamma_i} \right).
\]
\label{randef}
\end{definition}

\begin{definition}
A non-negative number \( R_i \) is an achievable rate for the message of sender \( i \) in a Byzantine multiple access classical-quantum channel \( \mathtt{W} \) under random coding if, for every \( \delta > 0 \), \( \epsilon > 0 \), and sufficiently large \( n \), there exists a random code such that  
\[
\frac{\log |M_i|}{n} > R_i - \delta,
\]  
and the average probability of decoding error for the message of sender \( i \) is less than \( \epsilon \) for any input sequence from the adversarial sender.

The closure of all achievable random rates for \( \mathtt{W} \) is called the random capacity region of \( \mathtt{W} \).
\end{definition}

The POVM \( \{ D_{m_i} : m_i \in M_i \} \) defines a channel \( C_{E_i} \), which maps

\[
\rho \rightarrow \sum_{m_i} \sqrt{D_{m_i}} \rho \sqrt{D_{m_i}}.
\]

\subsection{Key Differences to Classical Models}
\label{Keydiff}
    In the classical multiple access channel scenario, it is evident that the order in which the messages of the senders are decoded does not matter. In fact, most classical multiple access codes decode all messages simultaneously (cf.~\cite{Ahl4} and~\cite{Lia}). However, in the quantum multiple access channel scenario, the previously decoded operator can affect the subsequent decoded messages, as every measurement may alter the channel outputs.
    The standard solution to this issue in the quantum multiple access channel scenario is based on the following observation. Suppose the first decoding operator can decode the first message with a probability sufficiently close to 1, and assume this decoding operator is constructed using subspace projections. In this case, we can apply the Tender Operator Lemma (cf.~\cite{Og/Na} and~\cite{Win}) to show that this decoding operator does not significantly alter the channel outputs.

    The situation becomes significantly more subtle once one of the users may
    behave adversarially. In the Byzantine multiple access classical-quantum
    channel scenario, the Tender Operator argument can no longer be applied
    directly, since the receiver cannot guarantee that the first decoding operator
    succeeds with sufficiently high probability. Indeed, the first sender itself may
    be adversarial. Consequently, the corresponding decoding measurement may
    substantially disturb the channel output state. In this case, the remaining
    quantum state may no longer contain enough information to decode the second
    user's message reliably.
    
    \begin{figure}
        \centering
        \includegraphics[width=0.95\linewidth]{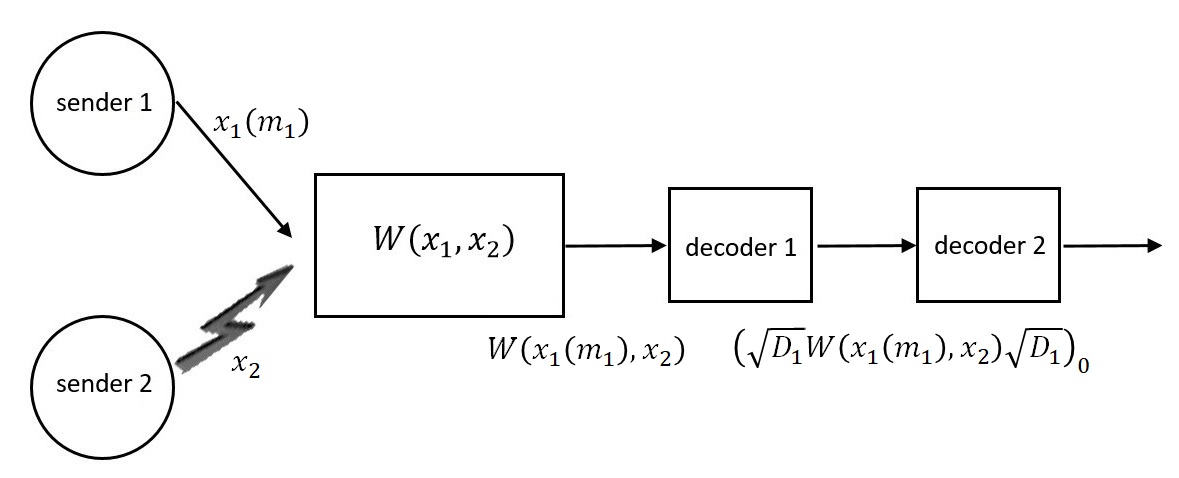}
        \caption{\label{2usermac}Sender \(1\) is trustworthy, while sender \(2\) is adversarial. The receiver first decodes the message from sender \(1\), followed by the message from sender \(2\).}
    \end{figure}
    \begin{figure}
        \centering
        \includegraphics[width=0.95\linewidth]{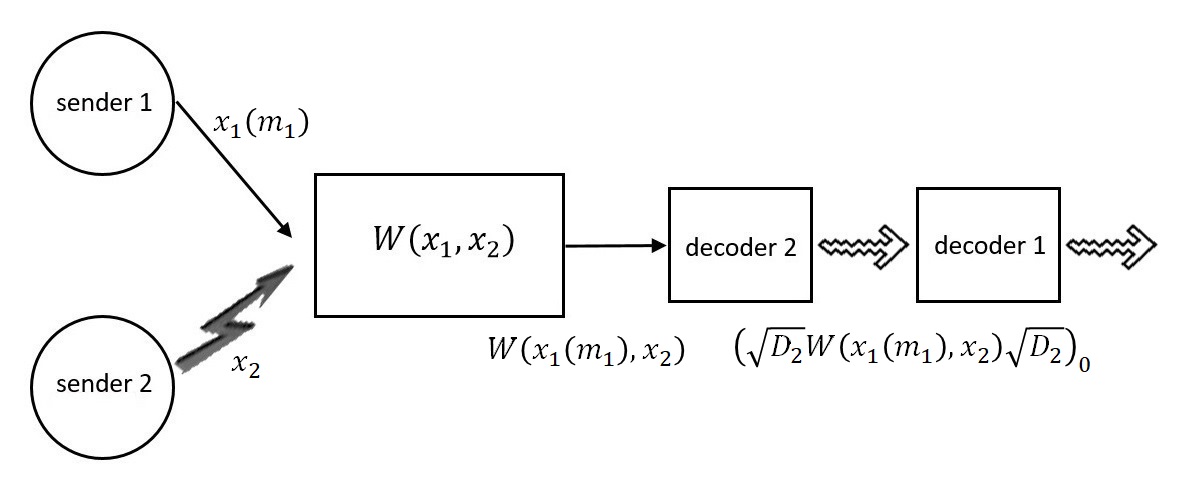}
        \caption{\label{2userbymac}\textit{Sender \(1\) is trustworthy, while sender \(2\) is adversarial. The receiver first decodes sender \(2\)'s message, followed by sender \(1\)'s message. In this case, the decoding operator for sender \(2\)'s message interferes with the decoding of sender \(1\)'s message.}}
    \end{figure}

    
    In~\cite{CaiQuantumByzantineMAC}, a simple one-shot example is presented showing that
    in the ``correct'' order,  namely 	decoding the message of
    the reliable users at first, the code will be reliable and error free.
    However, a ``bad
    order'', namely 	decoding the message of
    an  adversarial one at first, renders the code completely useless
    with the first POVM decoding operator. This
    is illustrated by \cref{2usermac,2userbymac}

    One may think that we can avoid problems caused by decoding order when
    we use simultaneous decoding. However, it is easy to see  that
    for our channel model, we have to employ a separable decoding operator for each sender's message, 
    rather than a simultaneous decoding approach.
     Simultaneous decoding is designed in such way that it
     decode all messages at the same time. It cannot
     decode the messages separately (else it would be
     separable decoding), thus
    simultaneous decoding only works when
 all senders can be trusted:
 Assume a $2$-user channel, 
a reliable simultaneous decoder $\{D_{m_1,m_2}: m_1\in M_1, m_2\in M_2 \}$ is designed in such way that it can only output a correct positive result when sender 1
sends $m_1$ and, at the same time,  sender 2 sends $m_2$.
When one of them is adversarial and does not send a
correct message, then the joint input will not
be in the domain of the decodable sequence.
Thus, the decoder operator $D_{m_1,m_2}$ can 
only tell if the joint message is indeed $(m_1,m_2)$
or not, it cannot check if only one of the two messages
is indeed $m_1$ or $m_2$.
This 
means that the decoder cannot decode the message 
of the  trustworthy sender correctly.
    In fact, the standard methods of quantum simultaneous decoding, applied in~\cite{Fa/Ha/Sa/Se/Wi}
    and~\cite{Se}, also utilize the same technique as the Tender Operator Lemma,
    which implies that these ``simultaneous decoding operators'' essentially decode one message at a time.
   

    Moreover, while the capacity of a classical 
    Byzantine multiple access channel can be achieved using a strong deterministic code, i.e.,
    without relying on any randomness, the same does not hold for the quantum results in our work (cf. \cref{randef}),
    as stated in \cref{ClvsQu}.
    

    
\subsection{Capacity Results and Techniques}
    Following the scheme of~\cite{Sa/Ba/De/Pr} and~\cite{Sa/Ba/De/Pr2},
    our idea for the  Byzantine multiple access coding is
    to combine the arbitrarily varying coding technique (scenario
    with an adversarial jammer with known identity) and the 
    multiple access coding technique (scenario
    with  trustworthy senders).
    
    The main achievability result of~\cite{CaiQuantumByzantineMAC} characterizes
    an achievable random coding rate region for Byzantine classical-quantum
    multiple access channels. In contrast to the classical setting, the achievable
    rates depend strongly on the decoding order. 
    
    
    More precisely, users decoded later may experience a rate penalty because
    previously applied decoding measurements can alter the remaining quantum state.
    The following theorem summarizes the corresponding achievable random coding
    region.
    
    \begin{theorem}[\cite{CaiQuantumByzantineMAC}]
    Let \( \mathtt{W} \) be a classical-quantum \( k \)-user Byzantine multiple access
    channel with at most one adversarial sender. Assume that the receiver decodes
    the messages in the order: sender \(1\), sender \(2\), \dots, sender \(k\).
    
    Then the random capacity region of \( \mathtt{W} \) satisfies
    \begin{align*}
    R_i \leq \min \Bigl(
    & \max_{p_i} \min_{p_j : j < i}
    I\bigl(p_i, E_j B \mid \{p_h : h \in \{1,\dots,i-1\}\setminus j\}\bigr), \\
    & \max_{p_i} \min_{p_j : j > i}
    I\bigl(p_i, B \mid \{p_h : h \in \{1,\dots,i-1\}\}\bigr)
    \Bigr).
    \end{align*}

where 
\(B\)  represents the outcome of \(  \mathtt{W}((p_1, p_2) \). and 
\( E_j B \) represents the outcome of \( C_{E_j} \circ \mathtt{W}(p_1,\cdots,p_k) \).    
    \label{lwback}
    \end{theorem}

    The asymmetry of the achievable rate region 
    in Theorem \ref{lwback} is a genuinely quantum phenomenon.
    Unlike in classical multiple access channels, we have to consider that  messages decoded later may
    be disadvantaged because previously applied POVM measurements. Consequently, even the two-user Byzantine
    multiple access channel, which appears relatively straightforward in the
    classical setting, can exhibit highly non-trivial behavior in the quantum
    scenario.
    Overall, the quantum extension of Byzantine communication reveals that many
    robustness concepts developed for classical systems depend implicitly on
    properties unique to classical information, such as non-invasive observation
    and deterministic verification. Once measurements become invasive and
    non-commutativity enters the model, substantially new adversarial phenomena
    emerge.

\section{Open Problems and Future Directions}
\label{sec:open_problems}

    Although significant progress has been achieved in recent years, Byzantine
    communication models still raise many fundamental open problems. The existing
    results already reveal rich connections between adversarial communication,
    distributed computation, authentication, and quantum information theory, but a
    general and unified theory is still far from complete. In this section, we
    outline several important directions for future research.
    
\subsection{Strong Deterministic Coding for Quantum Byzantine Channels}

    One of the most important open problems concerns deterministic coding for
    quantum Byzantine communication models (Cf. Section \ref{ClvsQu}). In the classical setting, deterministic
    coding theorems for arbitrarily varying channels and Byzantine multiple access
    channels can be established under suitable non-symmetrizability conditions~\cite{Ahl/Ca,Sa/Ba/De/Pr,Sa/Ba/De/Pr2} by the strong deterministic code concept, i.e., building a one-part code. In contrast, for arbitrarily varying
    classical-quantum channels, only random coding techniques,
    and therefore, only weak deterministic code techniques,
    are currently known
    \cite{Ahl/Bli}. Consequently, the presently available coding
    schemes for quantum Byzantine multiple access channels also rely fundamentally
    on shared randomness between the users and the receiver.
    
    A major open question is therefore whether strong deterministic coding theorems
    analogous to the classical AVC dichotomy can be established for quantum
    Byzantine MACs. Solving this problem would require substantially new
    derandomization techniques capable of handling quantum measurements,
    non-commutativity, and measurement disturbance simultaneously.

\subsection{General Multi-User Quantum Byzantine Networks}

    Existing quantum Byzantine communication results mainly focus on relatively
    simple multi-user settings, typically with two users and at most one
    adversarial participant~\cite{CaiQuantumByzantineMAC}. Extending these
    results to general multi-user quantum networks remains largely open.
    
    In particular, it is unclear how adversarial structures, collusion among
    adversarial users, and network topology interact in quantum communication
    systems. The situation becomes especially challenging when several decoding
    operations interact non-trivially due to quantum measurement disturbance.
    
    A comprehensive theory of quantum Byzantine network communication may require
    new tools from quantum network information theory, quantum hypothesis testing,
    and multipartite quantum state discrimination.
    
\subsection{Finite Blocklength Analysis}

    Most current results on Byzantine communication are asymptotic and characterize
    capacity regions in the limit of infinitely large blocklengths. However, many
    practical systems operate in regimes where only short packets and low latency
    are possible.
    
    Finite blocklength analysis for Byzantine communication models therefore
    constitutes an important open direction. In particular, one would like to
    understand:
    
    \begin{itemize}
    \item finite blocklength trade-offs between reliability and adversary
    identification,
    \item the effect of adversarial uncertainty on dispersion terms,
    \item robustness under short-packet communication constraints, and
    \item explicit coding constructions with practical decoding complexity.
    \end{itemize}
    
    Such results are especially relevant for future low-latency communication
    systems, including 6G architectures, distributed sensing, and autonomous
    control systems.

\subsection{Complexity of Robust Decoding}

    Most achievability proofs for Byzantine communication rely on typicality
    decoding, exhaustive consistency checks, or combinatorial verification
    arguments. While these methods establish information-theoretic achievability,
    their computational complexity is often prohibitively large.
    
    The development of efficient robust decoding algorithms therefore remains a
    major challenge. Important open questions include:
    
    \begin{itemize}
    \item complexity bounds for adversary identification,
    \item efficient decoding under colluding adversaries,
    \item practical coding constructions for Byzantine MACs, and
    \item efficient quantum decoding strategies for Byzantine quantum channels.
    \end{itemize}
    
    The computational complexity aspect is particularly important because Byzantine
    communication systems naturally arise in distributed architectures with
    resource-constrained devices.

\subsection{Learning-Based and Adaptive Adversaries}
\label{leandad}
    Most existing information-theoretic models assume adversaries with fixed
    strategies and fixed statistical knowledge. However, modern communication
    systems increasingly incorporate adaptive and learning-based mechanisms.
    
    This raises several new questions. How should Byzantine communication models be
    extended to adversaries that learn from previous transmissions? How does online
    adaptation affect symmetrizability and spoofability conditions? Can
    reinforcement learning or distributed learning systems remain robust against
    adversarial users that adapt dynamically to the protocol?
    
    These questions connect Byzantine communication directly to adversarial machine
    learning and distributed optimization.

\subsection{Robust Task-Oriented and Semantic Communication}
\label{robdis}
    Byzantine distributed function computation demonstrates that robustness depends
    not only on the communication channel but also on the specific task that the
    system aims to accomplish~\cite{AnilkumarSPNP25}.
    
    This observation suggests strong connections to semantic and task-oriented
    communication. Future communication systems may increasingly focus on recovering
    task-relevant information rather than reconstructing complete messages.
    
    Important open questions include:
    
    \begin{itemize}
    \item which semantic tasks remain robustly computable under adversarial
    uncertainty,
    \item how adversarial users affect semantic information measures,
    \item robustness of goal-oriented communication architectures, and
    \item integration of Byzantine robustness into semantic communication systems.
    \end{itemize}
    
    These problems are particularly relevant for distributed AI systems and
    autonomous multi-agent architectures.

\subsection{Integrated Sensing and Communication}

    Future communication systems are expected to tightly integrate communication,
    sensing, and distributed decision making. Byzantine robustness in such systems
    remains largely unexplored.
    
    In integrated sensing and communication (ISAC) systems, adversarial users may
    not only manipulate communication signals but also distort sensing
    measurements, environmental information, or state estimates. Consequently,
    communication robustness and sensing robustness become fundamentally coupled.
    
    Understanding these interactions may require entirely new information-theoretic
    frameworks combining adversarial communication, estimation theory, and
    distributed control.

\subsection{Unifying Classical and Quantum Robustness Theory}

    Finally, one of the most ambitious long-term goals is the development of a
    unified theory of robustness against internal adversaries that simultaneously
    covers classical communication, distributed computation, and quantum network
    communication.

    
    A deeper understanding of this common structure may eventually lead to a
    general theory of adversarial information processing across classical and
    quantum systems.





\section{Conclusion}
\label{sec:conclusion}
    Communication systems with adversarial users represent an emerging and rapidly
    growing area of multi-user information theory. In contrast to classical
    adversarial models, Byzantine communication considers internal adversaries,
    i.e., trustworthy users who may deviate arbitrarily from the prescribed
    protocol. This internal nature of the adversary substantially changes the
    communication problem and introduces new challenges that do not appear in
    classical channel models with external jammers.
    
    In this survey, we reviewed several recent developments on Byzantine
    communication models in both classical and quantum settings. We discussed
    reliable communication in Byzantine multiple access channels, communication
    with adversary identification, and Byzantine distributed function computation.
    Furthermore, we considered the extension to classical-quantum channels and
    highlighted several fundamental differences between classical and quantum
    robustness theory.

    A unifying feature of these models is that the same received observation may be consistent with several adversarial scenarios.
    Reliable communication or computation is possible only when the
    desired quantities remain invariant under all scenarios that are
    indistinguishable to the decoder. In classical Byzantine MACs, this principle
    appears through spoofability and generalized symmetrizability conditions. In
    distributed function computation, it leads to the notion of $s$-viability. In
    the quantum setting, the same phenomenon is complicated further by measurement
    disturbance and non-commutativity.
    
    The survey also highlights the evolving role of randomness in adversarial
    communication. While deterministic coding is often sufficient in classical
    systems under suitable structural conditions, quantum Byzantine communication
    currently relies fundamentally on random coding techniques. This reflects a
    deeper conceptual difference between classical and quantum robustness theory.
    
    Beyond their theoretical interest, Byzantine communication models are highly
    relevant for modern distributed communication systems. Emerging applications
    include wireless Internet-of-Things networks, distributed sensing, federated
    learning, autonomous multi-agent systems, semantic communication, and
    integrated sensing and communication architectures. In such systems, robustness
    against malicious participants is expected to become an increasingly important
    design criterion.
    
    At the same time, many important questions remain open. These include
    deterministic coding for quantum Byzantine channels, finite blocklength
    analysis, efficient robust decoding algorithms, adaptive adversarial models,
    and the development of robust semantic and task-oriented communication
    frameworks.
    
    Overall, Byzantine communication provides a rich information-theoretic
    framework for studying robustness against internal adversaries in distributed
    systems. The considered results already reveal deep connections between
    multi-user information theory, distributed computation, authentication,
    cryptography, arbitrarily varying channels, and quantum communication.
    Nevertheless, the field is still in its early stages, and many fundamental
    conceptual and technical challenges remain to be explored.


\bibliographystyle{splncs04}
\bibliography{references}

\end{document}